This paper is accepted for publication by IEEE Transactions on Medical Imaging, please cite as follow:

J. Chen, Y. Yuan. Decentralized Personalization for Federated Medical Image Segmentation via Gossip Contrastive Mutual Learning. IEEE Transactions of Medical Imaging, 2025, doi: 10.1109/TMI.2025.3549292

# Decentralized Personalization for Federated Medical Image Segmentation via Gossip Contrastive Mutual Learning

Jingyun Chen, and Yading Yuan, *Member, IEEE*

***Abstract*— Federated Learning (FL) presents a promising avenue for collaborative model training among medical centers, facilitating knowledge exchange without compromising data privacy. However, vanilla FL is prone to server failures and rarely achieves optimal performance on all participating sites due to heterogeneous data distributions among them. To overcome these challenges, we propose Gossip Contrastive Mutual Learning (GCML), a unified framework to optimize personalized models in a decentralized environment, where Gossip Protocol is employed for flexible and robust peer-to-peer communication. To make efficient and reliable knowledge exchange in each communication without the global knowledge across all the sites, we introduce deep contrast mutual learning (DCML), a simple yet effective scheme to encourage knowledge transfer between the incoming and local models through collaborative training on local data. By integrating DCML with other efforts to optimize site-specific models by leveraging useful information from peers, we evaluated the performance and efficiency of the proposed method on three publicly available datasets with different segmentation tasks. Our extensive experimental results show that the proposed GCML framework outperformed both centralized and decentralized FL methods with significantly reduced communication overhead, indicating its potential for real-world deployment. Upon the acceptance of manuscript, the code will be available at: https://github.com/CUMC-Yuan-Lab/GCML.**

## I. Introduction

In recent years, deep-learning models based on Convolutional Neural Network [1] and Vision Transformer [2] have demonstrated remarkable effectiveness in various medical image segmentation tasks. As the training of these models requires large amount of data, researchers often need to pool up data from multiple centers. While data pooling facilitates collaborative research and offers improved model performance, it often faces regulation obstacles such as Health Insurance Portability and Accountability Act (HIPAA) and General Data Protection Regulation (GDPR). To address this dilemma, Federated learning (FL)[3] was introduced in 2017 for collaborative model training without sharing privacy-sensitive data among centers. As patient privacy is a critical concern in medical field, FL has been applied to various areas of medical imaging, such as histopathology [4][5], image reconstruction [6][7], privacy preserving [8][9], semi-supervise and self-supervised learning [10][11]. Particularly, in a global-scale effort involving 71 institutions across six continents with over 6,000 patients, the model trained with FL paradigm substantially improved performance by 20% over the base model in the task of automated brain tumor segmentation on multi-parametric MRI scans, underlining the effectiveness of FL in real-world applications [12].

Although conventional FL showed great potential as a collaborative learning paradigm, it also has some limitations. First, it relies on a central server for model aggregation and distribution (hence referred as centralized FL), rendering it vulnerable to server failure and cyberattack during the operation. To eliminate this risk, decentralized FL was proposed where FL sites communicate to each other directly without a central server [13][14]. Another critical challenge comes from the problem of non-independent and identically distributed (non-IID) data among different sites. Because significant data heterogeneity is common among different medical centers due to the variations in patient population and clinical practice, it is difficult to train a global model to achieve optimal performance in all centers. It is therefore crucial to personalize the local model on each site for improved performance.

While various studies have been conducted to investigate novel approaches to address the above two challenges to achieve either decentralized or personalized FL [15][16], few attempts have been made to achieve both goals at the same time due to the following challenges. First, the peer-to-peer (P2P) communication in decentralized FL results in much less information exchange as compared to the client-server communication in centralized FL, because P2P

Jingyun Chen is with the Department of Radiation Oncology, Columbia University Irving Medical Center, United States (email: jc6171@cumc.columbia.edu).
Yading Yuan is with the Department of Radiation Oncology, Columbia University Irving Medical Center, United States (email: yading.yuan@columbia.edu).

communication happens between two sites at a time while client-server model usually involves multiple clients to exchange information with the server. As a result, decentralized personalization usually requires longer training time and/or more communication cost. Second, many existing methods for decentralized model exchange aims for averaging weights or aligning predictions of the incoming model and local models [17][18]. Such model exchange strategies may not guarantee optimal personalization, especially when the incoming model is trained on external site with distinctive data distribution. Third, to determine the sender-receiver mapping for model exchange at each round, some decentralized FL methods adopted a fixed communication scheme [18][19], thus requiring all the participating sites to maintain presence concurrently. Such communication scheme may become compromised if any site randomly drops in or drops out during the FL. Therefore, although some methods have been proposed, they either lack of effective strategies to learn a strong personalized model on local data with significantly reduced model exchange [17] or require pre-defined communication pattern that is not flexible and still prone to the failure of participating sites [18].

In this paper we propose a novel gossip contrastive mutual learning (GCML) strategy to address these limitations. GCML employs gossip protocol (GP) to allow participating sites to directly communicate to each other and exchange local models. To maximize the information usage from the incoming model to boost the performance of site-specific model, we generalize the Deep Mutual Learning (DML) [20] by introducing a contrastive learning mechanism such that both the incoming and local models learn to align to each other at voxels where the reference model correctly predicts their classes, while deviating away from the reference model when it provides wrong predictions. Since incoming models are trained on different sites where the datasets normally have different distributions, their performance on the local data can be highly heterogenous. Our deep contrastive mutual learning (DCML) provides a unified framework to handle this variation for a more efficient and robust knowledge exchange between the incoming and local models. Furthermore, considering that the target tumors are usually small compared with the whole image volume, applying contrastive mutual learning on the entire image may be unnecessary and suboptimal. Inspired by the overlap-based loss functions in image segmentation tasks, such as Dice Similarity Coefficient [21] or Jaccard Distance [22], we introduce a regional contrastive K-L Divergence ($rD_{CKL}$) to emphasize the difference between the prediction of two models at the tumor region for more effective mutual learning. Finally, to further harvest the unique insights derived from both the incoming and local models learned from their collaborative learning, we set the final local model as the fusion of these two models. We extensively evaluate and validate the proposed framework on three different medical image segmentation tasks.

The main contributions of this paper can be summarized as follow:

- We introduce a novel FL framework, among the first few attempts, to learn personalized models in decentralized environment based on each site's data characteristics.
- We develop a unified learning strategy to adapt the incoming model to the local data and contrastively balance between the local and incoming models towards improving the performance of the updated local model.
- On three publicly available datasets with different imaging modalities and segmentation tasks, our method achieve superior performance on testing sets over current state-of-the-art (SOTA) methods with significantly reduced communication cost.

## II. Related Work

### A. Centralized Federated Learning

In centralized FL, participating sites upload their model or model updates to an aggregation server, where a global model is synthesized and distributed back to each participating site. As the most established FL method, FedAvg [3] creates a global model on the server by taking the weighted average of all local models from participating sites. Since the conception of FedAvg, several modified methods have emerged with improved performance [23][24]. In 2018 Li et al. proposed FedProx to tackle the data heterogeny among FL sites [25]. By introducing a proximal term linking the global and local models, FedProx can stabilize the convergence behavior of FL in realistic heterogeneous networks. More recently, FedPIDAvg developed by Machler et al. [26] improved the aggregation function that considers not only the training sample sizes but also the loss decreases from the previous iteration as well as in integral term over past losses, which became the winning solution for the Federated Tumor Segmentation Challenge 2022 [27]. While centralized FL remains the most prevalent paradigm in federated learning, the reliance on a central server not only makes the entire system vulnerable to the server failure, but also precludes efficient usage of the improved computation capability in each participating medical center.

### B. Decentralized Federated Learning

Unlike centralized FL, decentralized FL does not depend on a central server for model exchange, instead it operates in a distributed manner using direct P2P communication [13], offering potential advantages in terms of computation cost under constrained communication [28], security [29][30] and model performance [19]. For example, Swarm Learning [14] allows the full circle of model aggregation and broadcasting in FedAvg to occur at a randomly selected client at each round. More recently, Gao et al extended the original Swarm Learning to handle data heterogeneity among FL sites by introducing a loss reflecting the label skewness of local sites [31].

Even though decentralized FL eliminated the reliance of central server, most aforementioned methods still rely on a "central" model to be used on all clients. Such a global model is unlikely to achieve optimal performance universally, due to various data distributions on each client. Therefore, instead of training a global model and distributing to all sites, it is more desirable to personalize the model for individual sites. This realization prompted the introduction of personalized FL methods, aimed at tailoring model updates to better suit the data nuances of each site.

### C. Personalized Federated Learning

Personalized FL allows the development of a local model tailored to the unique data distribution of each client, therefore can potentially improve model performance. In 2021 Li et al. introduced Ditto to align distinctive local models with the optimal global model [15]. Collins proposed FedRep to customize a classification head for each site [32]. Chen et al. introduced a retrogress-resilient framework to enhance client-specific model performance [33]. More recently, Wu et al. developed FEDORA with adaptive parameter propagation and selective regularization to mitigate negative model transfer [16]. Jiang et al. introduced IOP-FL framework to achieve personalized learning of clients both inside and outside the federation [34]. Despite of the improved performance, these methods adhere to a server-client structure as most of them require certain prior knowledge of model and/or data characteristics from all sites, which is best collected and stored by a central server.

Among few attempts to develop personalized models in a decentralized environment, BrainTorrent proposed by Roy et al conducts model aggregation on a randomly selected site rather than on central server [17] in each round. While the primary focus of this work was to develop a decentralized FL approach, model personalization was achieved by simply averaging the incoming models with the local model. In 2023, Kalra et al. introduced ProxyFL to conduct mutual learning between a proxy model and local models through server-free cyclic model transfer [18]. However, the mutual learning mechanism employed in ProxyFL aims to align the private (i.e. local) model with the proxy (i.e. incoming) model, which may not be optimal if the proxy model performs poorly on local data. Additionally, the site communication in ProxyFL is based on a predefined network of site adjacency, necessitating that all involved sites remain online throughout the FL process. This makes ProxyFL impractical in practical scenarios where sites may join or leave in the middle of FL. In this paper, we aim to rectify these limitations through our proposed method.

## III. Method

### A. Preliminaries

In this paper, the local data on $N$ sites are denoted as $[D_1, D_2, \dots, D_N]$. The corresponding local model weights are denoted as $[W_1, W_2, \dots, W_N]$. $P_R(Y|X)$ denotes the predicted probability distribution of model $R$ on the entire image, with input image $X$ and output classification label $Y$. Within the context of volumetric medical image segmentation, $P_R(y^{i,j,k}|x^{i,j,k})$ denotes the predicted probability distribution of model $R$ at voxel $(i, j, k)$ of image $X$, with input voxel $x^{i,j,k}$, and the output segmentation label $y^{i,j,k}$. The tumor ground truth at voxel $(i, j, k)$ is denoted as $g^{i,j,k} \in \{0,1\}$, with $g^{i,j,k} = 1$ for tumor and $g^{i,j,k} = 0$ for background. The tumor prediction value of mode $R$ at voxel $(i, j, k)$ is denoted as $q_R^{i,j,k} \in [0,1]$.

### B. Gossip Protocol

We use Gossip Protocol (GP) [35] as the decentralized communication protocol. In this protocol, each site periodically and randomly selects another site in the network to exchange information via a P2P communication. Given its distinct advantages in scalability, fault tolerance and adaptability, GP has been proposed as a decentralized alternative to server-client based FL [36][37]. However, compared to the universal model uploading and downloading in centralized FL, the P2P model transfer in vanilla GP FL framework provides much less information exchange. Furthermore, the random selection of model-exchanging pairs in GP leads to random sequence of incoming models to each receiver site. These factors make GP FL highly inefficient and unreliable. Therefore, we designed the following approaches to mitigate these limitations.

### C. Deep Mutual Learning

DML was firstly introduced by Zhang et al. [20] in 2D image classification task that allows two models to learn from each other in a collaborative manner thus to improve their performance collectively. The mutual learning was achieved by minimizing an additional loss term based on K-L Divergence ($D_{KL}$) between the probability distributions of two models' outputs. In particular, the $D_{KL}$ from model $A$ to model $B$ is defined as:

$$D_{KL}(P_B \| P_A) = \mathbb{E}_{Y\sim P_B} \log \frac{P_B(Y|X)}{P_A(Y|X)} \quad (1)$$

where $\mathbb{E}_{y\sim P_B}$ is the expectation of $P_B(Y|X)$.

DML has been introduced to FL for image classification [38][39][40] and language processing [41][42]. In this work, we firstly extended DML from image classification to 3D image segmentation where the K-L Divergence at each voxel, $D_{KL}^{i,j,k}$ , represents the difference between the predicted probability distributions of two segmentation models. In the context of decentralized FL where a model $S$ is sent from the sender site to the receiver site, let $R$ represent the segmentation model on the receiver site and the $D_{KL}$ from model $S$ to model $R$ is then defined as:

$$D_{KL}(P_R\|P_S) = \sum_{i,j,k} D_{KL}^{i,j,k}(P_R\|P_S) = \sum_{i,j,k} \mathbb{E}_{y^{i,j,k}\sim P_R} \log\frac{P_R(y^{i,j,k}|x^{i,j,k})}{P_S(y^{i,j,k}|x^{i,j,k})} \quad (2)$$

where $\Sigma_{i,j,k}$ represents the sum over all voxel indices $(i,j,k)$. For brevity, we write $x^{i,j,k}$ and $y^{i,j,k}$ as $x$ and $y$ respectively in the rest of this paper.

### D. Deep Contrast Mutual Learning (DCML)

In the original DML setting, the two models are initialized differently so they can learn distinct knowledge from the common data and transfer knowledge to each other by aligning their outputs. This process assumes that these two models perform similarly on the training data, ensuring alignment does not drive the models away from their own optimum of the loss function. While this assumption typically holds in both the original DML context [20] and in the centralized FL environment [38][39][40][41][42] due to the robustness of modern deep learning models to initialization, it may not apply in decentralized FL environments. Since the incoming model is trained on the sender site, its performance on the receiver dataset may suffer from the data heterogeneity between these two sites. As a result, simply aligning the incoming and local models may cause the better-performed model to move toward the opposite direction of its optimum, yielding suboptimal performance. To address this issue, we introduce contrastive learning mechanism to DML that enables two models to collaboratively learn from each other based on their performance on the local data.

To do this, we add a contrastive weight $c^{i,j,k} \in \{-1,1\}$ to the voxel-wise K-L Divergence $D_{KL}^{i,j,k}$ and define a contrastive K-L Divergence $D_{CKL}$ as:

$$D_{CKL}(P_R\|P_S) = \sum_{i,j,k} D_{CKL}^{i,j,k}(P_R\|P_S) = \sum_{i,j,k} \mathbb{E}_{y\sim P_R} \log\frac{P_R(y|x)}{P_S(y|x)}\, c_S^{i,j,k} \quad (3)$$

If model $S$ makes correct prediction at voxel $(i,j,k)$, we set $c_S^{i,j,k}$=1 so the minimization of Eq. (3) performs as the regular mutual learning that aligns the receiver model $R$ to the incoming sender model $S$. Otherwise, we $c_S^{i,j,k}$= -1 so the minimization of Eq. (3) essentially aims to enlarge the difference between the output of these two models, enabling model $R$ to contrastively learn from the failure of model $S$. This unified framework allows the receiver model $R$ continuously exchange information with the incoming model $S$, and dynamically adopts the useful information from $R$ to further enhance the receiver model performance.

### E. Regional Deep Contrast Mutual Learning (rDCML)

In Eq. (3), $D_{CKL}$ is computed on the entire image. Because tumor region usually occupies a small portion of the image volume, calculating $D_{CKL}$ on the entire image may not be able to catch the nuance difference between the prediction of two models at the tumor region, as majority of voxels belong to the high-confidence background region where the two models are prone to agree well. To encourage two models to align around the tumor area, we introduce regional contrastive K-L Divergence ($rD_{CKL}$). Given the tumor ground truth $G$, the $rD_{CKL}$ is defined as:

$$rD_{CKL}(P_R\|P_S,G) = \sum_{i,j,k} rD_{CKL}^{i,j,k}(P_R\|P_S,G) = \sum_{i,j,k} \mathbb{E}_{y\sim P_R} \log\frac{P_R(y|x)}{P_S(y|x)}\, c_S^{i,j,k} g^{i,j,k} \quad (4)$$

Besides true tumor, we are also interested in the regions predicted as tumor by the models. Therefore, we compute the $rD_{CKL}$ given the tumor prediction by model $R$ (denoted as $Q_R$):

$$rD_{CKL}(P_R\|P_S,Q_R) = \sum_{i,j,k} rD_{CKL}^{i,j,k}(P_R\|P_S,Q_R) = \sum_{i,j,k} \mathbb{E}_{y\sim P_R} \log\frac{P_R(y|x)}{P_S(y|x)}\, c_S^{i,j,k} q_R^{i,j,k} \quad (5)$$

Finally, the overall $rD_{CKL}$ from model $S$ to model $R$ can be defined as:

$$rD_{CKL}(P_R\|P_S) = \frac{rD_{CKL}(P_R\|P_S,G)+rD_{CKL}(P_R\|P_S,Q_R)}{\sum_{i,j,k} g^{i,j,k}+\sum_{i,j,k} q_R^{i,j,k}} \quad (6)$$

where $rD_{CKL}(P_R\|P_S,G)$ and $rD_{CKL}(P_R\|P_S,Q_R)$ are defined in Eq. (4) and (5) respectively. And the denominator $\sum_{i,j,k} g^{i,j,k} + \sum_{i,j,k} q_R^{i,j,k}$ is used for normalization.

### F. Gossip Contrastive Mutual Learning (GCML)

Finally, we introduce the unified GCML framework by integrating rDCML with GP communication, as described in Algorithm 1 and illustrated in Fig 1. Once a site receives an incoming model from a sender site, the two models, i.e., $S$ and $R$, are optimized alternatively and collaboratively using the local training data on the receiver site. In particular, by freezing the sender model weight of $W_S$, the receiver model weight can be updated by minimizing the following loss function:

$$L(W_R\|W_S) = (1-\lambda)JD(W_R) + \lambda rD_{CKL}(P_R\|P_S) \quad (7)$$

where $JD$ is a segmentation loss function based on Jaccard Distance [22], and $rD_{CKL}$ is the rDCML loss defined in Eq. (6), with $P_S$ and $P_R$ representing the output distributions of models $S$ and $R$, respectively. We use $\lambda$ to balance the two loss functions. In this study, we simply set it as 0.5 across all sites and datasets for consistency, but it should be noted that the optimal $\lambda$ may vary depending on the specific training dataset at each local site, which may be adjusted to further improve the performance.

Likewise, the total loss function to update the sender model weight $W_S$ can be defined as:

$$L(W_S \| W_R) = (1-\lambda)\, JD(W_S) + \lambda r D_{CKL}(P_S \| P_R) \quad (8)$$

Since the incoming model $S$ is also updated through DCML, we finally merge the updated model $S$ with the updated model $R$, to learn knowledge from both sides. This is achieved by the weighted average between the model parameters $W_R$ and $W_S$ after rDCML:

$$W_R = \frac{v_R W_R + v_S W_S}{v_R + v_S} \quad (9)$$

where $v_R$ and $v_s$ are the Jaccard Distance loss on the local validation data for model $R$ and $S$ respectively. We denote this procedure as the sender-receiver model merging. When a site receives multiple incoming models, it will process these incoming models sequentially. Meanwhile, different receiver sites can perform their local processing in parallel.

**Algorithm 1**. GCML

**Inputs**: initial model $W_n^0$ of site $n$, $n = 1, \dots N$
**Outputs**: final model $W_n^T$ of site $n$, $n = 1, \dots N$

**for** round $t = 1, \dots, T$ **do**:
  **for** $n = 1, \dots N$ **do**:
    Update local model on site $n$:
    $$W_n^t \leftarrow W_n^{t-1} - \eta \frac{\partial JD(W_n)}{\partial W_n^{t-1}}$$
  **end**
  Select pairs of sender and receiver sites:
  **for** each pair of sender and receiver sites **do**:
    **On sender site $S$:**
      Transfer incoming model $W_S^t$ to receiver site
    **On receiver site $R$:**
      Compute losses $L(W_R^t \| W_S^t)$ and $L(W_S^t \| W_R^t)$ with Eq. (7) and (8)
      Update incoming model and local model:
      $$W_R^t \leftarrow W_R^t - \eta \frac{\partial L(W_R^t \| W_S^t)}{\partial W_R^t}$$
      $$W_S^t \leftarrow W_S^t - \eta \frac{\partial L(W_S^t \| W_R^t)}{\partial W_S^t}$$
      Sender-receiver model merging with Eq. (9):
      $$W_R^t \leftarrow \frac{v_R W_R^t + v_S W_S^t}{v_R + v_S}$$
  **end**
**end**

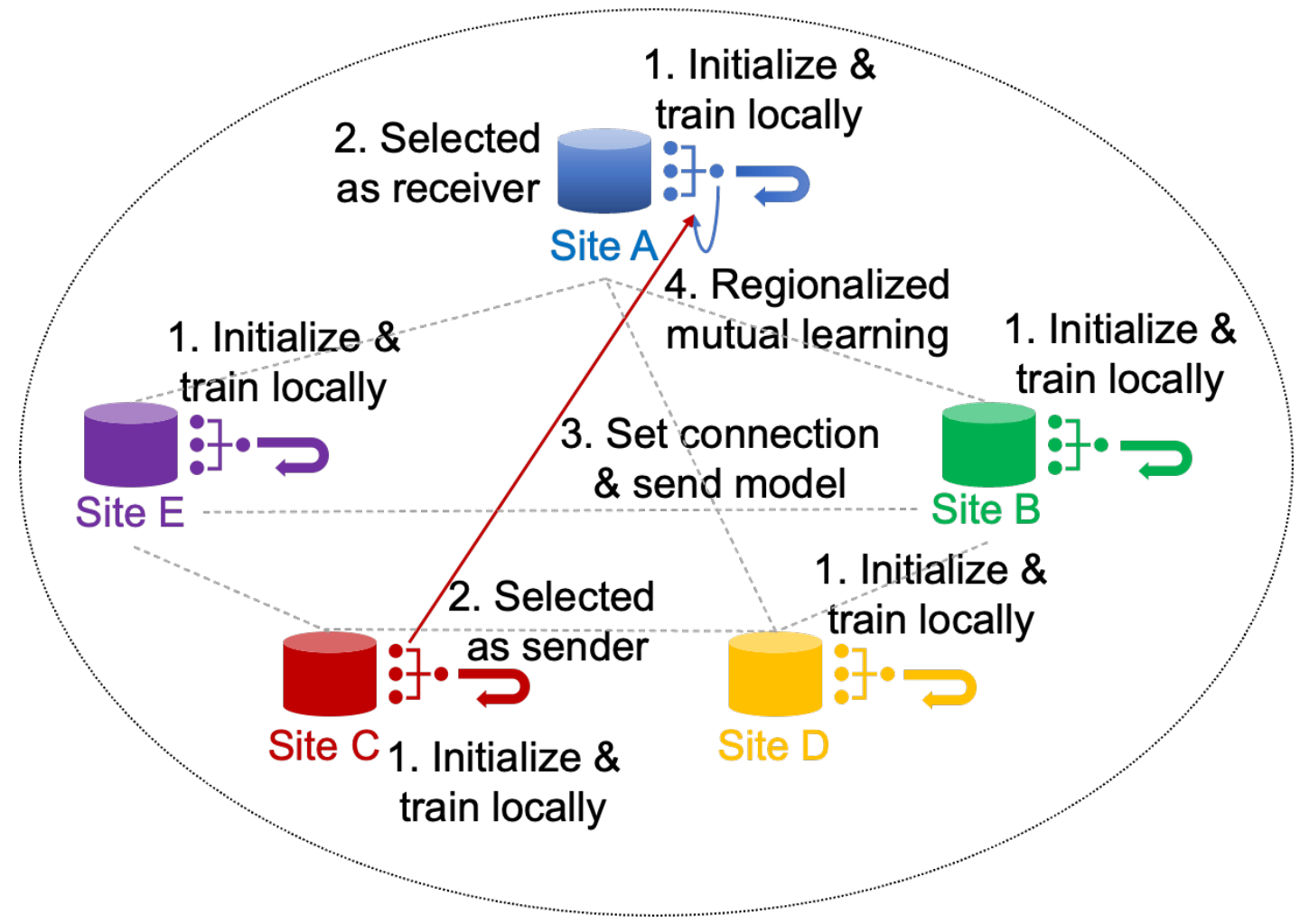

Fig 1. General scheme of GCML. After initialization and warm-up, pairs of sender and receiver are randomly selected at each round. Then the sender transfers its model to the corresponding receiver. And the receiver performs regionalized mutual learning between incoming model and its local model.

## G. Segmentation Model

We used Scale Attention Network (SANet) [43][44] as the automatic tumor segmentation model in this work. SANet features a dynamic scale attention mechanism to incorporate low-lever details with high-level semantics from feature maps at different scales, to better integrate information across different scales. It has demonstrated strong performance in different tasks such as head and neck tumor segmentation on PET/CT [45] and brain tumor segmentation on multi-parametric MRI images [46].

The architecture of SANet follows the classic encoding-decoding design (Fig 2a). The encoding pathway is built upon Residual Network [47]. To enhance the representative capability of the model, we added a squeeze-and-excitation model into each residual block to form a ResSE block (Fig 2c). The decoding pathway follows the reverse pattern as encoding one, but with a single ResSE block. The up-sampled feature maps are then added to the output of scale-attention block. Here we use summation instead of concatenation for information fusion. For the final loss function, we use deep supervision on each intermediate scale level, to regularize the model training and improve gradient flow propagation.

An example of Scale Attention block is showed in Fig 2b. This example is for the 2nd decoding scale level. The encoding outputs of other scales are first reshaped to the same dimension. Next, the features from all scales are summed up, followed by a global average pooling, and then the squeeze and excitation operation. Finally, the outputs of squeeze and excitation for different scales are normalized by a SoftMax and become a scale-specific weight vector for each channel. The final output of the scale attention block is the weighted sum of these vectors from different scales.

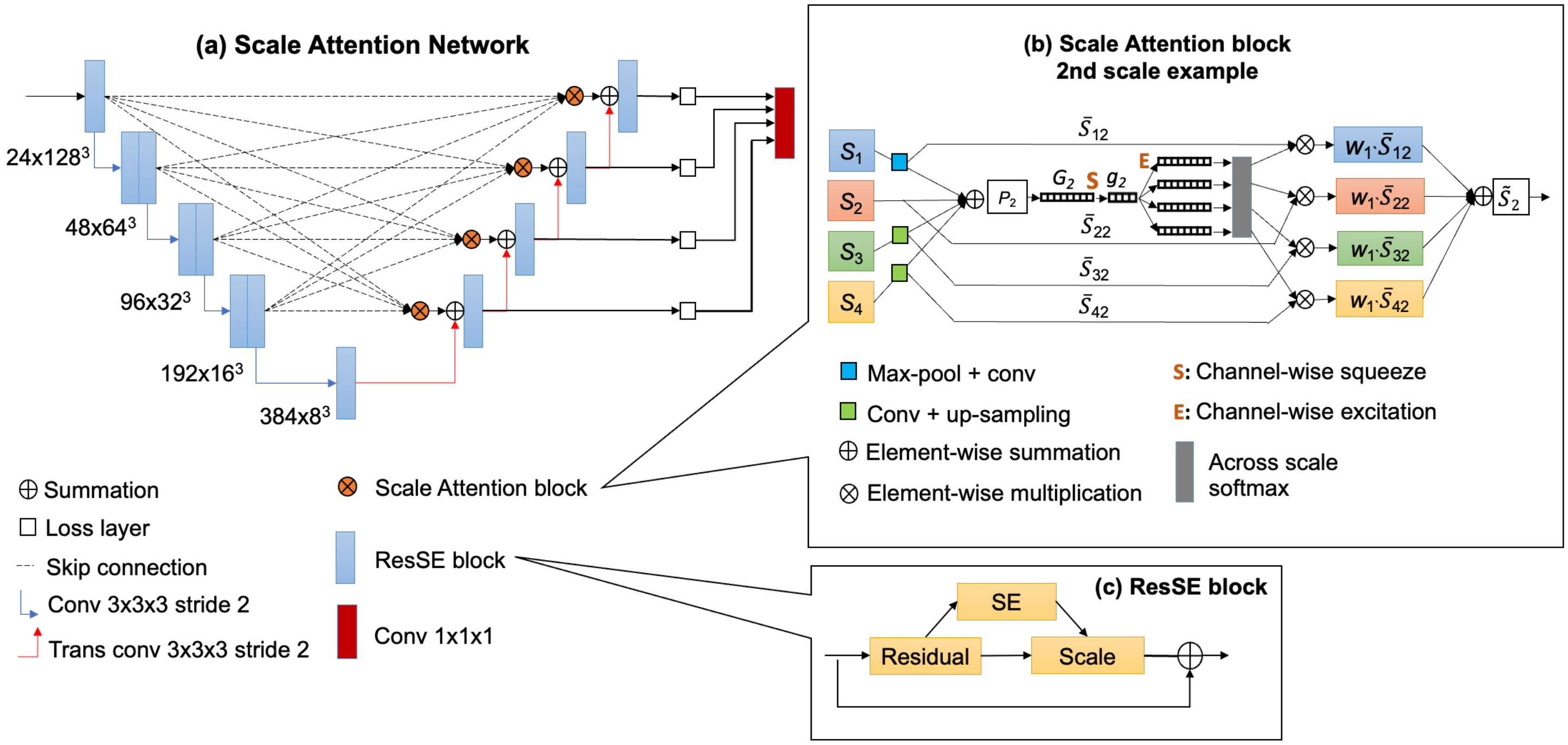

Fig 2. Architecture of SANet, including ResSE block and Scale Attention block

## IV. Experiments

### A. Data

We evaluated GCML on three publicly available datasets with different imaging modalities and segmentation targets, where all cases from the same dataset share the same imaging modalities. The first dataset is the training data of brain tumor segmentation challenge (BraTS) 2021 [48], which contains multi-parametric MRI sets including native T1-weighted, post-contrast T1-weighted, T2-weighted and a T2 Fluid-Attenuated Inversion Recovery (FLAIR) images. The BraTS data have three target regions, including enhanced tumor, necrotic and other non-enhancing tumor, and peritumoral edema. The second dataset is the training data of HEad and neCK TumOR segmentation (HECKTOR) 2020 challenge [49], which contains PET/CT images. The HECKTOR data has a single target region named gross tumor volume (GTV). The third dataset is the PanSeg data [50], a recently released dataset that contains the largest MRI pancreas dataset in the literature for automated pancreas segmentation. We used the T1-weighted image set in our study as it has more patients than the T2-weighted image set.

Since all three datasets provide hospital information regarding where the images were obtained, we use this information to formed individual sites for our FL experiments. For BraTS, we chose 8 representative sites where the number of cases ranges from 12 to 47. For HECKTOR, we used all the 200 training sets from four contributing sites. For PanSeg, we used the 384 T1-weighted MRI from 5 contributing sites. For each site in all three datasets, we randomly selected 70% of cases for training, 10% for validation, and 20% for local testing. The breakdown of case numbers in both datasets is shown in Table 1.

TABLE 1
CASE NUMBERS OF SITES IN TWO DATASETS

| Dataset | Site # | Total Cases | Training Cases | Validating Cases | Testing Cases |
|---|---|---|---|---|---|
| BraTS | 1 | 14 | 9 | 2 | 3 |
| | 2 | 30 | 21 | 3 | 6 |
| | 3 | 33 | 22 | 4 | 7 |
| | 4 | 35 | 23 | 4 | 8 |
| | 5 | 47 | 32 | 5 | 10 |
| | 6 | 22 | 15 | 3 | 4 |
| | 7 | 34 | 23 | 4 | 7 |
| | 8 | 12 | 7 | 2 | 3 |
| HECKTOR | 1 | 55 | 39 | 5 | 11 |
| | 2 | 18 | 11 | 3 | 4 |
| | 3 | 55 | 39 | 5 | 11 |
| | 4 | 72 | 51 | 7 | 14 |
| PanSeg | 1 | 161 | 113 | 16 | 32 |
| | 2 | 151 | 106 | 15 | 30 |
| | 3 | 30 | 21 | 3 | 6 |
| | 4 | 17 | 12 | 2 | 3 |
| | 5 | 25 | 17 | 3 | 5 |

### B. Baselines, SOTA Methods, and Implementations

We compared the performance of GCML with two baseline settings and five SOTA FL methods. In baseline settings, we firstly trained a Pool Model (**PM**) with centralized data sharing that pooled data from all sites; then we trained Individual Model (**IM**) on each site using its own local data without information exchange. The five FL methods include three centralized FL methods **FedAvg** [3], **FedProx** [25] and **FedPIDAvg** [26] [51], as well as two decentralized and personalized FL methods BrainTorrent (**BT**) [17] and **ProxyFL** [18]. All methods were implemented with PyTorch 2.0. On BraTS and HECKTOR datasets, each method was run for total 200 FL rounds without warmup. On PanSeg dataset, each method was run for total 500 FL rounds without warmup. By following the original publications, we set the proximal term scalar $\mu$ to 0.001 in FedProx [25], and the number of finetuning epochs to 2 in BT [17]. For ProxyFL, for simplicity we set both private and proxy model as SANet [43][44].

The experiments were conducted on 4 Nvidia GPUs. To ensure all sites participate model exchange as either sender or receiver, we randomly split the eight BraTS sites into four sender-receiver pairs, four HECKTOR sites into two pairs, and five PanSeg sites into three pairs (with one random sender site participating twice) at each FL round.

### C. Evaluation Metrics

We used Dice Similarity Coefficient (DSC) [21] between model prediction and manual outline (ground truth) as evaluation metric for model performance. Additionally, we also computed the 95% Hausdorff Distance (HD95) and Average Symmetric Surface Distance (ASSD) as evaluation metrics for segmentation accuracy. For the BraTS study, the mean DSC, HD95, and ASSD over three tumor subregions was computed for each case. For all three studies, the site-specific performance was evaluated on each site's own testing

TABLE 2
DSCs (%) (↑) FOR BASELINES AND FL METHODS

| Dataset | Site | Baselines | | FL Methods* | | | | | |
|---|---|---|---|---|---|---|---|---|---|
| | | PM | IM | FedAvg[3] | FedProx[25] | FedPIDAvg[26] | BT[17] | ProxyFL[18] | GCML(ours) |
| BraTS | 1 | 85.74 | 70.55 | 85.02 | 80.89 | 83.96 | 82.25 | 75.45 | **88.99** |
| | 2 | 85.27 | 81.62 | 84.51 | 83.66 | 84.26 | 84.3 | 82.34 | **84.69** |
| | 3 | 96.16 | 95.32 | **95.49** | 94.86 | 94.82 | 94.28 | 93.78 | 94.85 |
| | 4 | 95.22 | 92.54 | 94.41 | 93.44 | 94.23 | 94.10 | 92.05 | **94.63** |
| | 5 | 95.64 | 91.35 | **94.56** | 94.00 | 93.89 | 93.77 | 91.03 | 93.92 |
| | 6 | 92.30 | 87.83 | 90.81 | 90.83 | 87.14 | 89.85 | 90.85 | **91.37** |
| | 7 | 89.98 | 87.39 | 88.75 | 89.10 | 89.15 | **89.48** | 87.76 | 89.15 |
| | 8 | 94.49 | 89.91 | 94.02 | 94.17 | 94.67 | 93.75 | 92.07 | **94.90** |
| | Avg. | 92.55 | 88.65 | 91.62 | 90.95 | 91.05 | 91.04 | 89.11 | **91.87** |
| HECKTOR | 1 | 85.85 | 84.30 | 84.52 | 84.52 | 83.41 | 82.3 | 85.03 | **85.52** |
| | 2 | 79.68 | 67.89 | **79.68** | 74.47 | 79.20 | 79.27 | 70.29 | 78.00 |
| | 3 | 69.71 | 70.34 | 67.94 | **73.54** | 72.53 | 67.52 | 70.07 | 71.92 |
| | 4 | 79.34 | 73.20 | 77.13 | 77.14 | 76.66 | 73.81 | 76.22 | **79.10** |
| | Avg. | 78.52 | 74.93 | 76.89 | 77.91 | 77.64 | 74.96 | 76.36 | **78.78** |
| PanSeg | 1 | 84.35 | 81.70 | 82.62 | 82.46 | 83.48 | 81.75 | 82.22 | **84.05** |
| | 2 | 82.56 | 78.07 | 79.29 | 77.50 | 80.85 | 77.88 | 79.56 | **82.37** |
| | 3 | 86.02 | 81.93 | 84.19 | 83.35 | 84.48 | 81.69 | 74.30 | **86.71** |
| | 4 | 80.40 | 55.99 | 78.88 | 77.66 | 78.99 | 77.61 | 58.19 | **81.92** |
| | 5 | 79.87 | 70.61 | 81.09 | 80.00 | 80.63 | **81.82** | 73.66 | 80.41 |
| | Avg. | 83.32 | 78.54 | 81.18 | 80.22 | 82.15 | 80.06 | 79.03 | **83.27** |

*The highest DSCs among FL methods are bolded

data, while the overall performance was calculated as the average of site-specific performances weighted by the number of testing cases on each site. For IM, BT, ProxyFL and GCML, the local models were evaluated for site-specific performance, but we used the global model in PM, FedAvg, FedProx, and FedPIDAvg, to evaluate its performance on the local data as they do not have local models.

## V. Results

This section contains the main experiment results. In part A, we evaluated the performances of GCML, baselines and other FL methods on BraTS, HECKTOR and PanSeg datasets. In part B, C, and D, since BraTS dataset has more participating sites, we focused on this dataset to demonstrate the robustness of the proposed GCML to the sequence of incoming models and conduct the ablation studies to evaluate the effectiveness of each key component of the GCML framework. In part E, we evaluated the overall performance of SA-Net for pancreas segmentation by comparing its performance with PanSegNet [50] and nnUNet v2 [52] on the PanSeg dataset. All the reported results were based on the model performance on the testing data.

### A. Comparison with Baselines and FL Methods

Table 2 lists the segmentation performance of different approaches in terms of DSC, while Table 3 for HD95 and ASSD. On BraTS data, PM baseline achieved the best overall DSC as it was trained with centralized data pooling. By utilizing data from different sites, all FL methods improved the DSC as compared to the IM baseline, with GCML achieving the best performance among FL methods. For site-specific performance, GCML outperformed IM baseline on 7 out of 8 sites, with improvement ranging from 2%-18%. The only exception is Site 3, where GCML had slightly lower DSC than IM. The high DSC of IM (95.32%) suggested that Site 3 may already have sufficient local data to train its own model, therefore had little further gain from FL. The performance improvement by GCML can also be observed on the segmentation result of individual cases. One example is shown in Fig 3 (upper three rows), where GCML correctly eliminated both false positive and false negative appeared on

TABLE 3
HD95 / ASSD (MM) (↓) FOR BASELINES AND FL METHODS, LOWER VALUES REPRESENT BETTER PERFORMANCES

| Dataset | Site | Baselines | | FL Methods* | | | | | |
|---|---|---|---|---|---|---|---|---|---|
| | | PM | IM | FedAvg | FedProx | FedPIDAvg | BT | ProxyFL | GCML |
| BraTS | 1 | 4.73/1.67 | 22.87/12.01 | 4.67/1.81 | 13.86/4.85 | 5.46/2.22 | 4.86/1.85 | 19.40/8.81 | **3.20/1.05** |
| | 2 | 5.17/3.31 | 13.07/7.30 | 5.91/2.96 | 9.26/5.87 | 6.61/**2.16** | 8.33/5.46 | 9.45/5.08 | **4.62**/2.24 |
| | 3 | 2.12/0.76 | 8.45/2.03 | 2.28/0.84 | 2.52/1.14 | 4.05/0.98 | **2.12/0.83** | 2.41/1.04 | 4.90/1.33 |
| | 4 | 2.41/0.60 | 4.60/1.11 | 2.60/0.71 | 3.15/1.03 | 2.59/0.75 | 2.49/0.67 | 3.36/0.92 | **2.12/0.59** |
| | 5 | 2.44/0.67 | 4.15/1.53 | 2.75/0.80 | 3.17/1.03 | **2.72/0.80** | 2.72/0.89 | 3.58/1.39 | 2.97/0.93 |
| | 6 | 2.16/0.75 | 3.59/1.18 | 3.50/0.97 | 4.34/0.97 | **2.16/0.81** | 2.79/1.03 | 2.79/1.03 | 2.34/0.84 |
| | 7 | 5.75/1.51 | 11.66/3.34 | 6.75/2.13 | 6.45/2.10 | 6.63/2.10 | 8.17/**1.82** | 9.97/3.01 | **5.93**/1.83 |
| | 8 | 2.83/0.98 | 8.76/2.80 | 2.85/0.99 | 3.24/1.20 | 3.34/1.16 | 2.89/1.01 | 4.66/1.91 | **2.77/0.88** |
| | Avg. | 3.36/1.21 | 8.48/3.22 | 3.82/1.34 | 5.08/2.05 | 4.11/1.29 | 4.24/1.63 | 6.03/2.42 | **3.70/1.22** |
| HECKTOR | 1 | 3.10/0.95 | 4.86/1.26 | 3.25/1.08 | **3.05/1.00** | 3.49/1.28 | 4.97/1.94 | 3.90/1.06 | 4.61/1.14 |
| | 2 | 3.70/1.49 | 14.26/5.77 | **3.76/1.48** | 3.76/1.53 | 3.82/1.53 | 4.24/1.83 | 9.01/2.63 | 3.77/1.62 |
| | 3 | 5.54/2.79 | 5.26/2.59 | 6.92/3.49 | 8.00/3.34 | 8.35/2.71 | 14.97/5.24 | 9.72/3.25 | **6.04/2.63** |
| | 4 | 4.88/1.58 | 6.49/3.10 | 5.25/1.96 | 5.38/2.00 | 5.48/2.08 | 5.11/**1.73** | 8.24/2.81 | **5.08**/1.82 |
| | Avg. | 4.45/1.73 | 6.48/2.72 | **5.01**/2.09 | 5.30/2.05 | 5.56/1.98 | 7.70/2.76 | 7.53/2.43 | 5.08/**1.84** |
| PanSeg | 1 | 6.16/1.38 | 7.96/1.96 | 7.82/1.94 | 7.67/1.84 | **6.22**/1.46 | 7.93/2.06 | 6.74/1.60 | 6.33/**1.44** |
| | 2 | 8.15/2.21 | 11.72/4.05 | 11.59/4.33 | 13.51/5.79 | 9.53/2.73 | 11.82/4.81 | 11.26/4.28 | **7.68/1.98** |
| | 3 | 3.83/1.39 | 5.59/1.68 | 4.26/1.55 | 4.92/1.70 | 4.41/1.56 | 8.03/2.96 | 15.45/6.23 | **3.89/1.29** |
| | 4 | 8.86/1.95 | 22.03/7.76 | 8.29/2.11 | 8.05/2.10 | 7.80/2.06 | 8.52/2.24 | 18.14/5.77 | **7.38/1.62** |
| | 5 | 6.99/1.75 | 15.77/4.41 | **5.88/1.60** | 6.93/1.75 | 6.53/1.72 | 6.29/1.78 | 12.42/3.18 | 6.96/1.81 |
| | Avg. | 6.92/1.75 | 10.33/3.15 | 8.92/2.84 | 9.72/3.39 | 7.47/2.01 | 9.39/3.21 | 10.04/3.29 | **6.75/1.67** |

*The lowest HD95 and ASSD among FL methods are bolded

the IM results. The HD95 and ASSD results are consistent with DSC, GCML method showed lower average HD95 and ASSD than IM and other FL methods, confirming the improved performance.

On HECKTOR data, GCML outperformed both baselines and other FL methods on DSC. This result clearly demonstrated the advantage of personalized model through GCML. Across all individual sites, GCML consistently surpassed the performance of IM baseline, with improvement ranging from 1%-10%. Among the other FL methods, only FedProx consistently outperformed IM baselines on all 4 sites, yet with less improvement than GCML. The remaining FL methods (FedAvg, FedPIDAvg, BT and ProxyFL) showed improvement over IM on some sites but not all. As compared to other FL methods, GCML outperformed FedAvg, FedProx, FedPIDAvg and BT on 3 out of 4 sites. Notably, GCML outperformed ProxyFL on all 4 sites, indicating the advantage of DCML over DML in the decentralized environment with reduced information exchange. The HD95 and ASSD results (Table 3) are generally consistent with DSC, except for the HD95 and ASSD of PM, and the HD95 of FedAvg, which are lower than those of GCML. One representative case is showed in Fig 3 (middle three rows), where GCML avoided the over-segmentation in IM and other FL methods, generating the most accurate prediction.

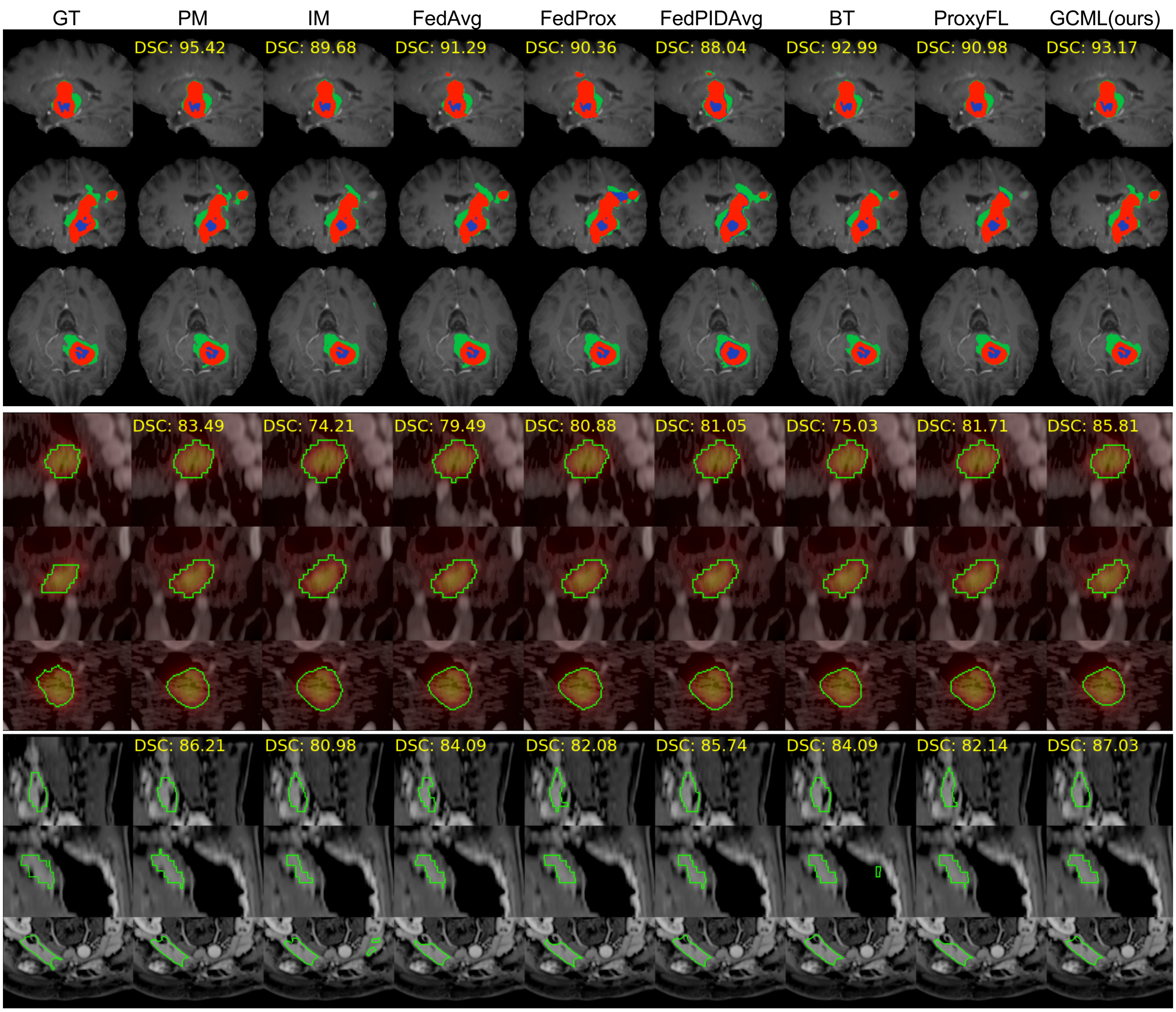

Fig 3. Compare of ground truth (GT) and predicted segmentations by different models. Rows 1-3 correspond to a BraTS case, with T1ce scan as background and target regions of enhanced tumor (red), necrotic and non-enhanced tumor (blue), and edema (green). Rows 3-6 correspond to a HECKTOR case, PET heatmap overlaid on CT scan as background, and gross tumor outlined in green. Rows 7-9 correspond to a PanSeg case, with T1 scan as background and pancreas outline in green. The DSC of each model is printed in yellow on top row of each case.

TABLE 4
THE MEAN AND STANDARD DEVIATION (S.D.) OF DSC (%) FOR GCML OVER 10 TRIALS

| Site | Trial 1 | Trial 2 | Trial 3 | Trial 4 | Trial 5 | Trial 6 | Trial 7 | Trial 8 | Trial 9 | Trial 10 | Mean ± S.D. |
|---|---|---|---|---|---|---|---|---|---|---|---|
| 1 | 88.99 | 82.60 | 85.91 | 82.35 | 89.69 | 80.15 | 85.28 | 88.05 | 84.58 | 89.87 | 85.75 ± 3.38 |
| 2 | 84.69 | 84.82 | 84.62 | 83.46 | 83.93 | 84.46 | 84.38 | 83.34 | 84.14 | 84.39 | 84.22 ± 0.51 |
| 3 | 94.85 | 94.74 | 95.32 | 95.76 | 95.90 | 95.67 | 95.78 | 95.99 | 95.74 | 95.68 | 95.54 ± 0.43 |
| 4 | 94.63 | 94.83 | 94.16 | 94.50 | 94.17 | 94.59 | 93.96 | 94.55 | 93.97 | 94.36 | 94.37 ± 0.30 |
| 5 | 93.92 | 94.54 | 94.56 | 93.76 | 94.22 | 94.16 | 94.50 | 94.41 | 93.87 | 93.02 | 94.10 ± 0.48 |
| 6 | 91.37 | 90.28 | 92.24 | 91.53 | 91.06 | 91.11 | 91.90 | 91.79 | 89.25 | 91.47 | 91.20 ± 0.87 |
| 7 | 89.15 | 89.18 | 89.60 | 90.57 | 89.54 | 89.01 | 89.22 | 88.44 | 89.07 | 88.77 | 89.26 ± 0.57 |
| 8 | 94.90 | 94.52 | 94.60 | 94.51 | 94.74 | 95.11 | 94.86 | 94.51 | 94.15 | 94.40 | 94.63 ± 0.28 |
| Avg. | **91.87** | **91.52** | **91.91** | **91.57** | **91.97** | **91.42** | **91.79** | **91.80** | **91.30** | **91.69** | **91.68 ± 0.22** |

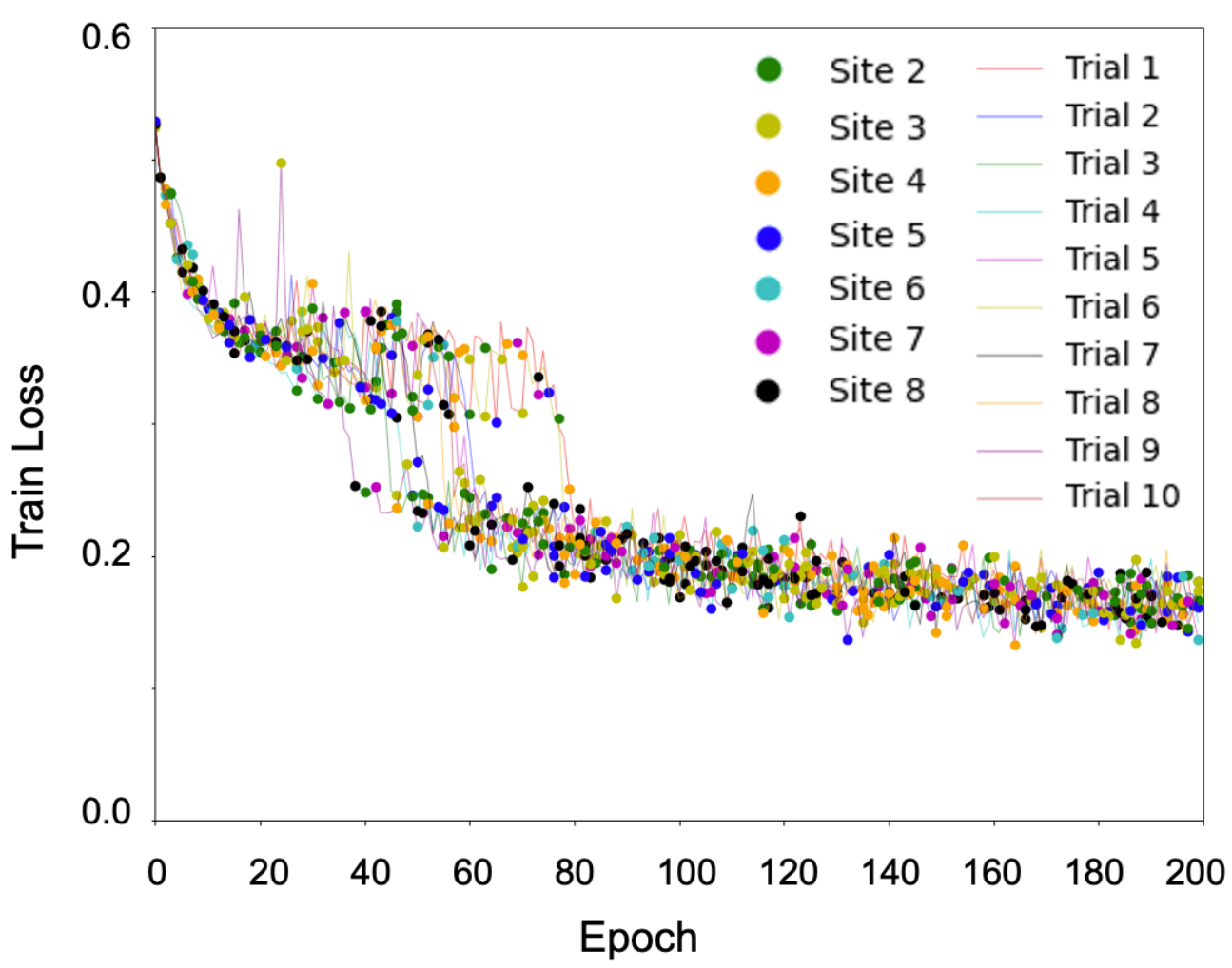

Fig 4. Training losses of Site 1 under different sequences of incoming models, driven by different random seeds

On PanSeg data, PM achieved highest DSC. GCML outperformed IM and all other FL methods on 4 out of 5 local sites, as well as on average. Notably, the local DSC improvements with GCML compared to IM are more pronounced on smaller sites (e.g., Site 3-5, with fewer than 50 local cases, showing DSC increases of 5% to 26%) than larger sites (e.g., Site 1 and 2, with more than 100 local cases, showing increases of only 2% to 4%). This indicates the smaller sites with less local data can potentially benefit more from GCML, by exchanging models with larger sites. On HD95 and ASSD, PM achieved the lowest values. Our GCML outperformed IM and all other FL methods on 4 out of 5 local sites, as well as on average. One representative case is showed in Fig 3 (bottom three rows), where GCML avoided the under-segmentation in FedAvg, FedProx, ProxyFL, as well as the false positive in IM and BT, resulting a higher DSC than other methods.

### *B. Effects of Incoming Model Sequence*

Learning order is generally considered as important to sequential optimization problems such as decentralized GP FL since it determines when and where a participant can receive models. We evaluated the effect of incoming model orders, i.e. the sequence of incoming models, on the final FL performance by conducting 10 trials of GCML on BraTS data. In reach trial, we used different random seeds to create different sequences of sender-receiver pairs while keeping the rest of settings consistent.

Fig 4 shows the training loss of different sequences of incoming models on Site 1 over the 10 trails. Here different sender sites are marked as dots of different colors, while sites belonging to the same sequence (i.e. trial) are linked with line segments of same color. The figure illustrated a notable variance among incoming model sequences across the 10 trials. However, GCML consistently achieves a stable final performance.

Table 4 presents the site-specific and overall DSC of GCML across the 10 trials. The findings reveal that GCML consistently exhibits minimal deviation in overall performance (standard deviation of 0.22) across these trials. This stability owes much to the contrastive learning mechanism within GCML, enabling local sites to effectively adapt to diverse data distributions presented by incoming models. Both the tabulated results and Fig 4 underscore the robustness of GCML against the random sequencing of incoming models.

### *C. Ablation Study*

To evaluate the contribution of each key component to the final GCML performance, we progressively added the following components to the base algorithm that only included DML: 1) contrastive learning; 2) regionalization; and 3) sender-receiver merging. Additionally, we also evaluated the information exchange strategy that directly merges the sender and receiver models without mutual learning i.e. merging only.

As illustrated in Table 5, each of the proposed components contributed to the performance improvement. In particular, contrastive learning resulted in a noteworthy 2.26% improvement in overall performance, underlining the effectiveness of rDCML in the decentralized FL environment with significantly reduced information exchange. Furthermore, the merging of the updated sender and receiver models after rDCML introduced an additional 2.14% improvement, which represented the additional contribution from the incoming model through model fusion. As showed in Eq. (7) and (8), both incoming model and local model are updated through rDCML, resulting in two personalized models, both of which can contain unique insights of local data. Therefore, the merging of local model with incoming model after rDCML provides further improved performance. In comparison, merging the original sender and receiver models without rDCML led to a lower overall DSC of 88.74%.

### D. Robustness with Different Data Splits

In our study, we intentionally included both small and large sites to represent the diverse data sizes across different sites in real-world settings. Collaborative studies among medical centers in practice encourage participation from small centers, such as community hospitals. These small centers usually have unique but limited local data, which not only enables them to contribute diverse data to the collaborative federated learning, but also allows them to benefit the most from the stronger models trained on large sites. In our previous experiments, we followed the common practice in FL studies [12] to randomly allocate 20% of cases for testing and reported the performance of different methods based on this single data split in Table 2 and 3.

TABLE 5
DSCs (%) UNDER DIFFERENT GCML CONFIGURATIONS

| Site | Merging Only | DML Only | DML + Contrast | DML + Contrast + Region | DML + Contrast + Region + Merging |
|---|---|---|---|---|---|
| 1 | 70.63 | 73.69 | 73.28 | 75.68 | 88.99 |
| 2 | 82.07 | 82.83 | 83.30 | 83.70 | 84.69 |
| 3 | 94.33 | 92.63 | 94.39 | 94.79 | 94.85 |
| 4 | 92.75 | 89.84 | 93.71 | 93.80 | 94.63 |
| 5 | 93.23 | 90.12 | 93.45 | 93.03 | 93.92 |
| 6 | 89.60 | 88.20 | 88.54 | 89.58 | 91.37 |
| 7 | 85.29 | 84.95 | 88.81 | 86.61 | 89.15 |
| 8 | 88.44 | 86.42 | 87.05 | 89.71 | 94.90 |
| Avg. | **88.74** | **87.36** | **89.62** | **89.73** | **91.87** |

DML = Deep Mutual Learning
Contrast = Contrastive Learning
Region = Regionalization
Merging = Sender-receiver Model Merging

To evaluate how different data splits affect the performance of various methods, we compared GCML with other privacy-preserved approaches using five-fold cross-testing. In each fold, we reserved non-overlapping 20% of cases for testing while using the rest 70% of cases for model training and 10% for validation. In this way, all the cases on one site will be used to test the model's performance.

Fig 5 plots the DSC values of all the cases obtained from the five-fold cross-testing. Here the y axis is the DSC values from GCML, and x axis represents the other privacy-preserved approach to be compared. It clearly shows that most points lie in the region above the diagonal line, indicating GCML achieved better performance than its counterpart.

The violin plots in Fig 6 show the DSC distribution obtained by different methods. In addition to the average DSC from the three different brain tumor subregions (WT: whole tumor; TC: tumor core; ET: enhanced tumor), we plot the DSC distribution of individual subregions. As compared to other methods, the DSC values of GCML are more concentrated in the higher value range.

We also evaluated the statistical significance of the difference between GCML and other privacy-preserved methods. Since the Shapiro-Wilk test [52] indicated none of the DSC data followed a normal distribution, we used Wilcoxon Rank Sum test [53] to evaluate statistical significance. Our results show that GCML achieved statistically significant improvement (at significance level $\alpha$ =0.05) than all the other methods in the WT region, and statistically better than most of the other methods in the rest regions. Our findings demonstrate that GCML consistently achieved better performance than the other privacy-preserved approaches with different data splits.

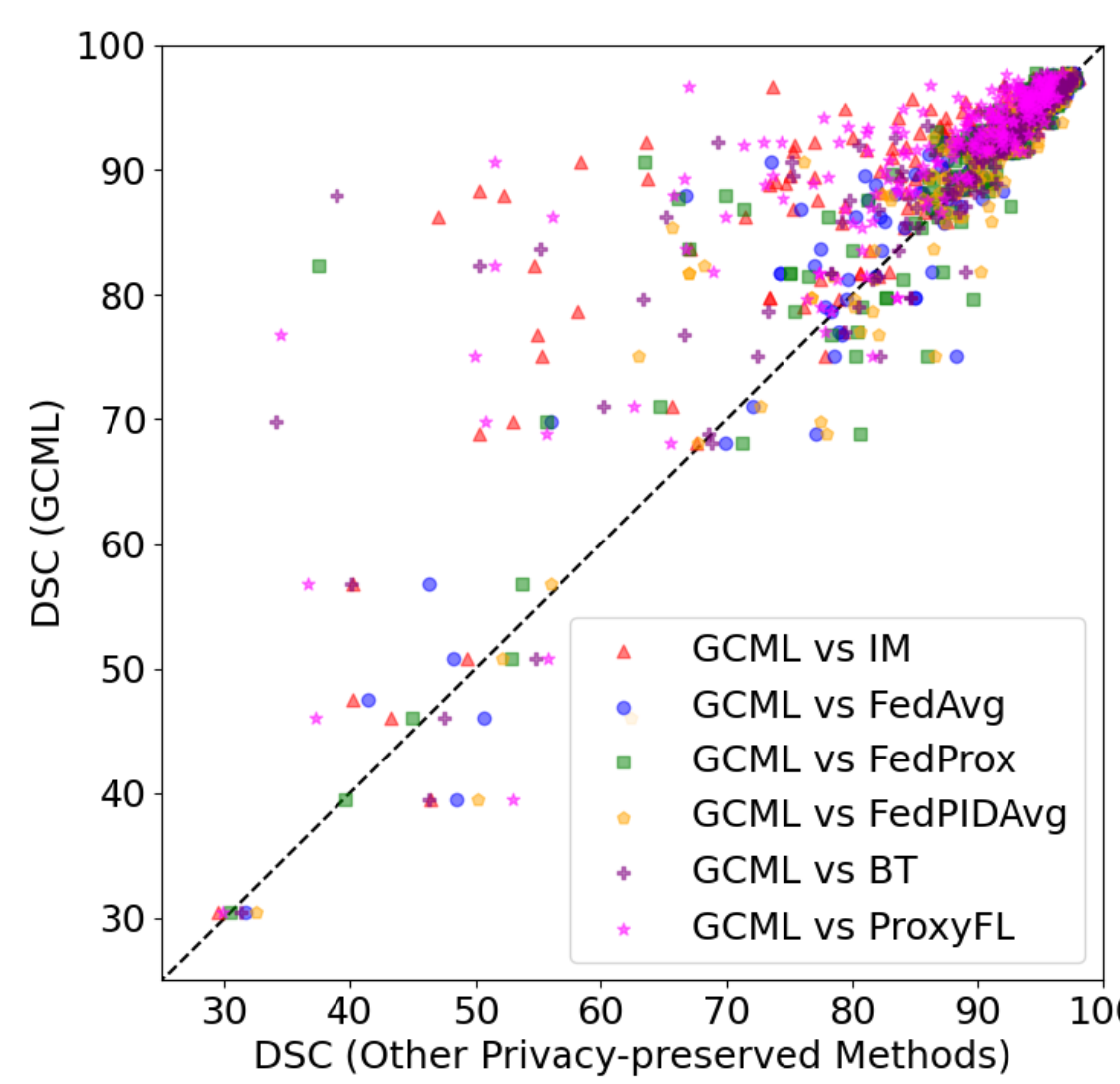

Fig 5. Scatter plot of avg. DSC between GCML vs other privacy-preserved learning methods (IM, FedAvg, FedProx, FedPIDAvg, BT, and ProxyFL). The DSC is averaged over WT, TC, and ET regions. For each point, its Y coordinate is the avg. DSC of GCML, while its X coordinate is the avg. DSC of the privacy-preserved method to be compared.

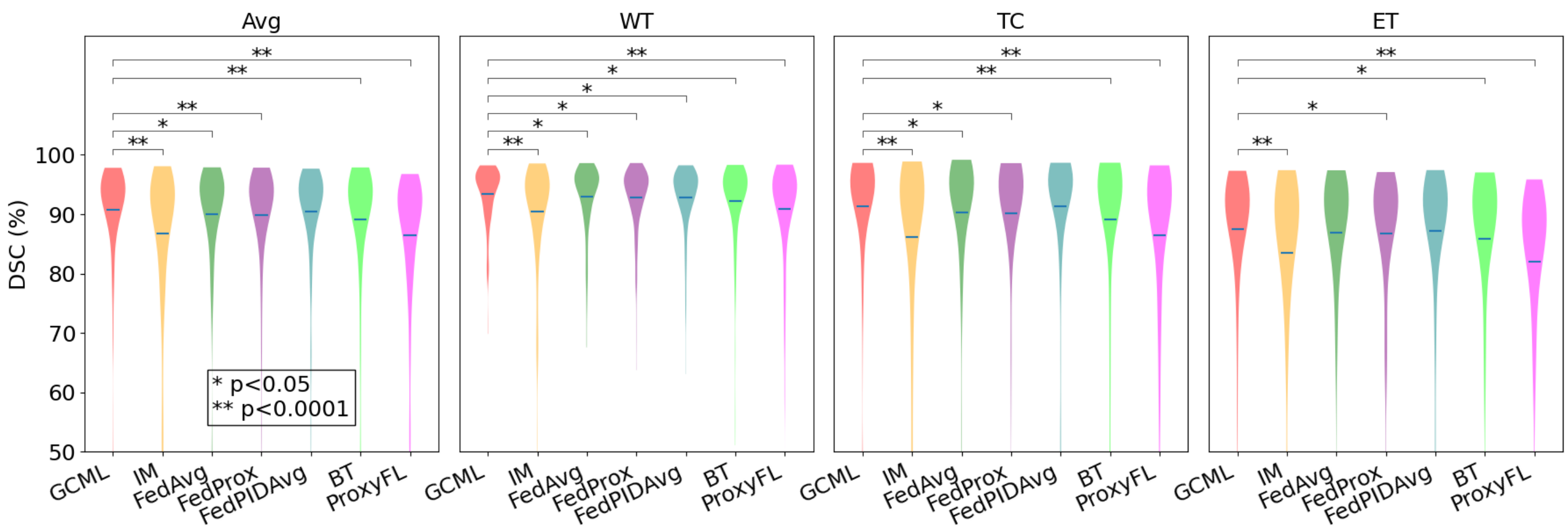

Fig 6. Violin plots of DSC for difference privacy-preserved learning methods (GCML, IM, FedAvg, FedProx, FedPIDAvg, BT, and ProxyFL) over different tumor regions (Total, WT, TC, and ET, where Avg is the average of WT, TC and ET). The horizontal bars within the violin plots represent the mean values. The group difference significance between GCML and other privacy-preserved learning methods are marked above the violin plots, with * for p<0.05 and ** for p<0.0001.

### E. Overall Performance of SANet on PanSeg Data

In addition to the performance in the FL environment, it is also important to understand the overall performance of a model on a particular task to ensure it is suitable for this task. SA-Net has been validated on both BraTS21 and HECTOR datasets in our previous studies [45][46]. Therefore, we evaluated its performance on the PanSeg dataset as MR imaging brings unique challenges, such as motion artifacts, field inhomogeneities and largely anisotropic voxel sizes, in segmenting pancreas on 3D images. We followed the same experimental setup as the original paper [50]: the combined data from Site 1 and 2 were used for model training and 5-fold cross validation, and the 72 patients from the rest three sites were used as external testing cases. We also compared SA-Net with nnUNet model (version 2) [54]. To ensure a fair model comparison, we applied nnUNet's preprocessing pipeline to the raw image data so both SA-Net and nnUNet use the same preprocessed data as input.

Table 6 compares the segmentation performance of different models (PanSegNet results are copied from Table 4 of [50]). With only 10.3M model parameters trained with 500 epochs, SANet outperformed both PanSegNet and nnUNet on both internal validation and external testing, highlighting its superior generalization capability. As compared to nnUNet that ran 1000 epochs to train 31.1 M parameters, SA-Net achieves this performance with about one-third of model parameters and 50% of training epochs. This efficiency is crucial for both federated learning and clinical deployment as it enables faster training convergency while requiring less computational resources and model complexity to achieve

TABLE 6
COMPARE OF DSC (%), HD95(MM) AND ASSD (MM) FOR SANET AND PANSEGNET. THE ROWS WITH * INDICATE THE PERFORMANCE IN THE TRAINING SITES USING A 5-FOLD CROSS-VALIDATION (INTERNAL VALIDATION), WHILE THE REST OF THE ROWS PRESENT THE PERFORMANCE IN THE TESTING SITES (EXTERNAL TESTING).

| | DSC (%) (↑) | | | HD95(mm) (↓) | | | ASSD (mm) (↓) | | |
|---|---|---|---|---|---|---|---|---|---|
| Site | SANet | PanSegNet | nnUNet | SANet | PanSegNet | nnUNet | SANet | PanSegNet | nnUNet |
| 1* | **84.77** | 83.70 | 83.39 | **5.66** | 6.79 | 7.44 | **1.32** | 1.42 | 2.10 |
| 2* | 85.99 | **86.44** | 86.35 | 6.63 | **5.91** | 6.03 | 1.45 | **1.22** | 1.51 |
| Internal Avg. (1 and 2)* | **85.35** | 85.02 | 84.82 | **6.13** | 6.37 | 6.76 | 1.38 | **1.32** | 1.82 |
| 3 | **84.88** | 81.55 | 82.94 | **4.49** | 5.64 | 4.70 | **1.50** | 1.80 | 1.64 |
| 4 | 79.72 | **79.80** | 77.10 | 8.30 | **8.01** | 9.51 | 1.86 | **1.67** | 2.33 |
| 5 | **78.97** | 76.21 | 77.50 | **6.94** | 14.26 | 8.65 | **1.99** | 2.98 | 2.52 |
| External Avg. (3,4, and 5) | **81.61** | 79.27 | 79.67 | **6.24** | 9.23 | 7.21 | **1.75** | 2.19 | 2.10 |

The highest DSC and lowest HD95 and ASSD are bolded high segmentation performance. Our results confirm that SA-Net is a promising solution for accurate and efficient pancreas segmentation on MRI images.

## VI. Discussion

While centralized federated learning with a global model was a natural choice when FL was firstly introduced for collaborative learning with mobile devices [3], its application in medical domain has raised concerns such as its vulnerability to the failure of model server and the limited capability to handle diverse data distributions of participating sites. Therefore, decentralized FL and personalized FL have been introduced to address these limitations. Specifically, image segmentation is a highly personalized practice as it not only depends on the site-specific data characteristics such as imaging devices, data acquisition protocols and patient demographics, but also is substantially impacted by tumor delineation protocols and the level of clinical expertise at the site.

This paper presents a unified framework, GCML, to learn personalized models in a decentralized FL environment. As compared to the methods proposed in previous studies, a distinct feature of GCML is that each participant maintains a local model and information exchange takes place between two participants in a P2P manner under Gossip Protocol. An advantage of GP is that it doesn't require all participating sites to stay online for all FL rounds. If a site goes offline, it won't disrupt random pairings in subsequent rounds. However, the original GP can lead to high communication cost due to the inefficient information exchange and random communication. To address this, we introduced Deep Contrastive Mutual Learning to enhance the quality of information exchanged during each GP communication. As shown in the ablative study, DCML led to 2.26% performance improvement. Furthermore, DCML serves as a separatable module to GP in our proposed framework: GP is used to select sender and receiver sites, while DCML is employed for pairwise model exchange. Therefore, the overall complexity of the system remains manageable.

While robust and flexible, this decentralized FL environment requires effective knowledge exchange in each communication without the general knowledge across all the participating sites. By recognizing that the primary motivation for each participant in joining FL is the aspiration to improve the segmentation performance of their local model through the collective expertise of other participants, we introduce DCML. This approach, though simple, proves highly effective in transferring knowledge between the incoming model and the local model through collaborative training on local data. Since DCML allows the receiver to selectively adopt the incoming models based on their performance on the receiving site, our framework was shown to be robust to the randomness of incoming model sequences, as demonstrated in Table 4. As a result, our method achieves high performance while significantly reducing the communication frequency, thus lowering the overall communication overhead in FL, as shown in Table 7.

Since the original DML was introduced to FL framework in [38], several extensions of DML have been developed for the FL environment. These include incorporating regularization terms between local and global models [39], adding a mean square error cost for hidden states [41][42], and using cross-correlation learning prior to DML [40]. The proposed GCML differs from these methods in three key aspects. First, while prior methods primarily focused on aligning the incoming and local models, GCML incorporates a contrastive learning mechanism into DML, allowing the local model to deviate from the incoming reference model when it produces incorrect predictions. Second, while previous works applied DML to image classification or language processing tasks, we extended DML to image segmentation by introducing voxel-wise K-L Divergence between the two models. Finally, whereas prior works utilized DML within a centralized FL environment based on a server-client structure, our GCML method is decentralized and does not rely on an aggregation server.

Data variation across different sites plays an important role in federated learning. Our study employed three datasets with different imaging modalities and segmentation targets. The variations in patient population, image acquisition protocols and ground truth identification have been discussed in the original publications [48][49][50]. The column (a) of Fig 7 illustrates the distribution differences of input data across various sites using t-distributed stochastic

TABLE 7

COMMUNICATION OVERHEAD OF DIFFERENT METHODS ON DIFFERENT DATASETS, IN MEGABYTES (MB)

| Dataset | Baselines | | FL Methods* | | | | | |
|---|---|---|---|---|---|---|---|---|
| | PM | IM | FedAvg | FedProx | FedPIDAvg | BT | ProxyFL | GCML (ours) |
| BraTS | 0.00 | 0.00 | 66.00 | 66.00 | 66.00 | 28.87 | 33.00 | **16.50** |
| HECKTOR | 0.00 | 0.00 | 32.99 | 32.99 | 32.99 | 12.37 | 16.49 | **8.25** |
| PanSeg | 0.00 | 0.00 | 103.13 | 103.13 | 103.13 | 41.25 | 51.56 | **30.94** |

*The lowest communication bits among FL methods are bolded

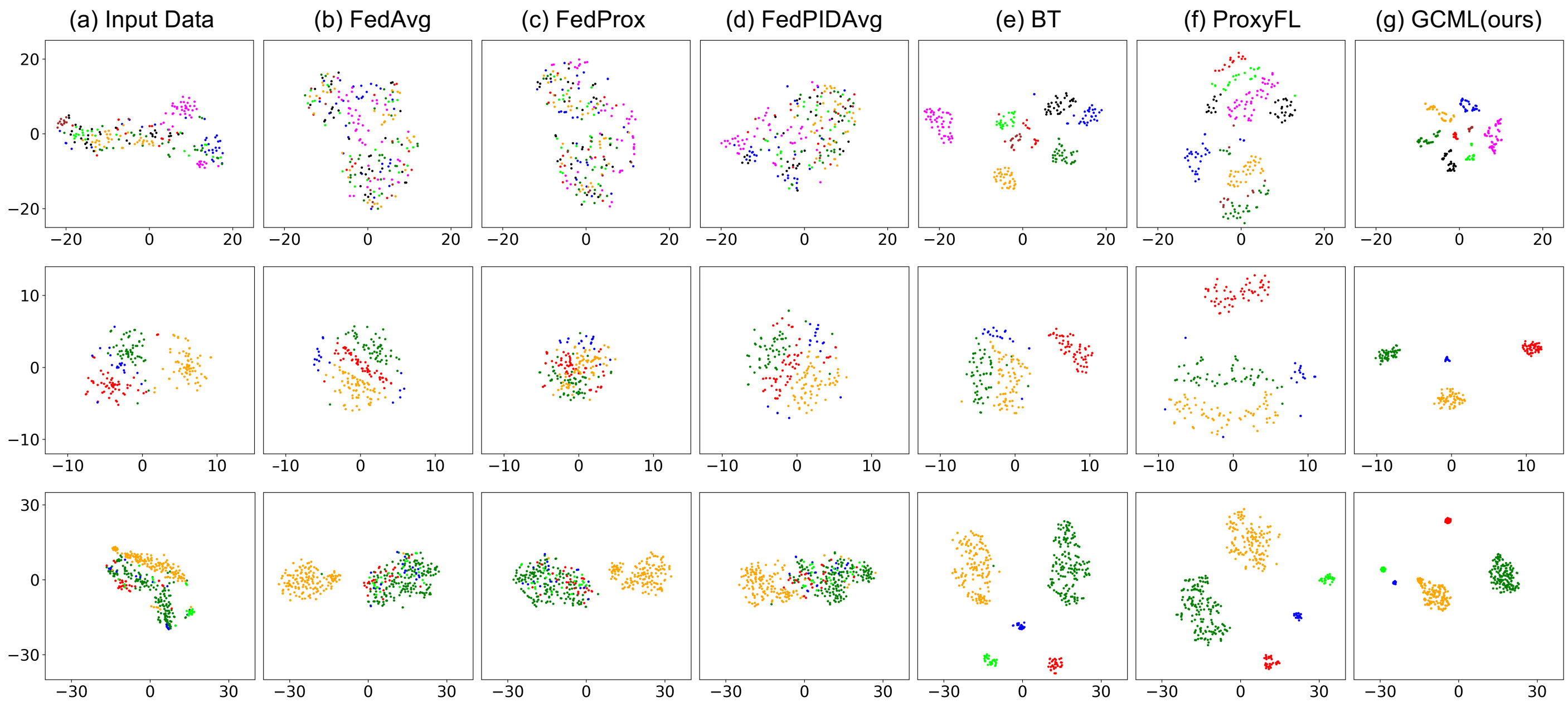

Fig 7. t-SNE visualization of input data distributions (a) and learned features from different FL methods (b-g). The rows from top to bottom corresponds to BraTS, HECKTOR and PanSeg datasets respectively. Each data point represents a case. Different colors represent different sites.

neighbor embedding (t-SNE)[55][56]. It clearly shows that the majority of sites form a unique distribution, especially for HECKTOR and PanSeg datasets. The more overlapping among BraTS sites is partially due to the standardized pre-processing steps on the raw BraTS data before being released to the public [48].

To further understand how these distribution differences affect model performance, we used t-SNE to visualize the distribution of learned features across various FL methods. As shown in columns b-g of Fig 7, there are distinct differences in how centralized and decentralized FL methods handle inter-site variance. Centralized FL methods (FedAvg, FedProx and FedPIDAvg) aim to mitigate inter-site differences by learning a shared global representation. Therefore, their learned features exhibit a mixed distribution across different sites, with a higher degree of mixture indicating better model generalization. While these methods can learn invariant representations when data distribution is relatively homogeneous among different sites (e.g., BraTS), it becomes more challenging for them when each site presents a more unique distribution. For example, in the PanSeg dataset, the features learned by FedAvg and FedProx still exhibit distinct clustering patterns, indicating inferior performance compared to FedPIDAvg.

In contrast, decentralized FL methods (BT, ProxyFL and GCML) explicitly preserve local variations by allowing models to adapt to site-specific data distributions, resulting in site-specific clusters in feature space. As non-IID data distribution makes a global model unlikely optimal for each individual FL site, it is crucial to train a personalized model on each site while leveraging information from other sites through FL. Across all three datasets, GCML consistently learned features that form distinct and compact clusters in feature space, highlighting its superior ability to capture inter-site differences while maintaining high predictive performance. As demonstrated in Table 2 and 3, GCML achieved better performance than other FL methods. The improvement is more prominent on HECTOR and PanSeg datasets where each site presents more distinct data distribution.

It should be noted that contrastive learning has been widely used in self-supervised learning as an effective framework for learning feature representations that aims to increase the distance between the representations of different images, i.e., *negative pairs*, while reducing the distance between the representations of different augmented views of the same image, i.e., *positive pairs* [57][58]. Inspired by this principle, DCML is proposed to amplify the difference of model outputs at voxels where the reference model fails to yield correct predictions (*negative pairs*), and conversely, diminish the discrepancy at voxels where the reference model performs correctly (*positive pairs*). As compared to the traditional contrastive learning, DCML has the following distinct characteristics. Firstly, DCML is designed for supervised learning in a decentralized FL setting while contrastive learning is for self-supervised learning. Supervised contrastive learning has demonstrated improved performance and robustness in various machine learning tasks due to the effective usage of label information [59]. DCML extends supervised contrastive learning such that voxels are grouped into either *negative* or *positive pairs* based on the performance of the reference model, instead of the classes they belong to. Secondly, contrastive learning typically performs clustering in the intermediate feature embedding space such that preferable features can be learned for downstream tasks. However, in a voxel-wise classification problem such as image segmentation, effectively representing a voxel in the feature embedding space is still an open question. Meanwhile, Zhang et al. in [20] points out that aligning intermediate feature map would diminish the cohort diversity and harm the ability of mutual learning. Therefore, DCML focuses on output-oriented knowledge distillation by minimizing a contrastive loss based on the final predictions between models. This also makes DCML different from other contrast learning based FL approaches such as MOON [60].

Communication cost is an important consideration when implementing FL method in a real world. As model sizes grow to enhance performance, the associated increase in communication overhead can extend the duration of each FL round. In real-world FL scenarios characterized by limited communication bandwidth and waiting time, FL methods with high communication efficiency offer a significant advantage. To compare the communication cost of GCML, baselines and other FL methods, we computed the communication bits, i.e., the model weights transferred in bits, per round [61] (Table 7). PM trained on centralized data, and IM trained on individual sites without model sharing, therefore neither of them requires model transfer. For the 8-site BraTS study, GCML had only 4 model transfers per round when sending models from the senders to the receivers. In comparison, the communication bits nearly doubled for BT, doubled for ProxyFL, and quadrupled for both FedAvg and FedProx. Th ratios are similar in the 4-site HECKTOR study and 5-site PanSeg study, with GCML showing the lowest cost. These results show that GCML is a highly efficient learning strategy that achieves better performance with much less communication.

The communication overhead in GCML is directly controlled by the number of sender/receiver pairs, which can be easily adjusted thanks to the flexibility of the Gossip Protocol. In general, more sender-receiver pairs lead to more frequent model exchanges and better final performance. However, increasing exchanging pairs will also increase the communication overhead, presenting a trade-off between model performance and communication efficiency. In comparison, the communication scheme in ProxyFL is dependent on the number of sites. While this setting protects ProxyFL from the random effect of incoming model sequence, it also requires all the sites to announce their availability before FL (so that they can be registered on the mapping) and stay available throughout the FL (otherwise the mapping is broken). These requirements make ProxyFL less flexible to the dynamic change of participating sites.
In this study, we employed 4 pairs of sender-receiver for the 8-site BraTS study, 2 pairs of sender-receiver for the 4-site HECKTOR study, and 3 pairs of sender-receiver for the 5-site PanSeg study. The rationale is to allow all sites to participate in FL at each round, either as the sender or the receiver. To explore the performance of GCML under different settings of sender-receiver pairs, we further experimented on the 8-site BraTS study with 1 and 2 pairs of sender-receiver, to compare with the original setting of 4 pairs (Table 8). As expected, the overall performance of GCML improved as the number of pairs increased.

There is often a trade-off between learning time and quality in FL. The proposed GCML belongs to personalized federated learning where the local model exchanges information with the incoming model through mutual learning [18][38][39][40][41][42]. Therefore, it is expected that a personalized FL requires additional training time than non-personalized ones, i.e., FedAvg, where the incoming model is directly used as the local model. For example, ProxyFL, a personalized FL framework, on average required 121% more training time than FedAvg in our experiments, but since GCML contrastively learns between incoming and local models for a more efficient and effective information exchange, it only required 49% more training time than FedAvg to achieve supervisor performance.

GCML does not require the receiver to receive multiple models from other sites in each FL round. The number of incoming models in each FL round can be set by each client, so the client would know the maximum number of models it will receive in each FL round. Since contrastive mutual learning is conducted between the local model and each of the incoming models sequentially, the local training can start once it receives the first incoming model while waiting for others to come. From practical consideration, a maximum waiting time can be set so the site will stop waiting if no model arrives in the specified time period to prevent the site from waiting indefinitely. Since GCML is a decentralized FL framework where each site performs independently except during model exchange, these local models are not expected to be the same so they can fit the unique data distribution at different sites, removing the need for global synchronization and potentially improving scalability [62]. This is different from the traditional applications of the Gossip Protocol in distributed systems where all nodes should eventually converge to the same state.

The need of warmup in federated learning (FL) remains a topic of debate [63][64]. On one hand, warmup allows local models to converge before being shared with other sites. On the other hand, warmup delays the actual model-exchanging rounds, which may hinder model optimization. To investigate the impact of warmup rounds on GCML performance, we compared GCML's performance with 100, 50 warmup rounds, and the original setting without warmup (Table 9). While 50-round warmup achieved the highest overall performance, 0-round warmup allows for more communication rounds without significant loss in performance. In a real-world environment, 0-round warmup holds further advantages, as it allows the newly joined sites to start model exchange without delay.

One limitation of our study is that we simulated the FL environment on a single workstation with multiple GPUs. This constraint prevented us from including some established FL methods like Swarm Learning [14] as a baseline. Because the main difference between Swarm Learning and FedAvg is the former method conducts model aggregation on random clients instead of central site. However, when both server and clients are simulated on the same workstation, Swarm Learning loses its distinctive feature and essentially becomes equivalent to FedAvg. This simulation study also makes some metric for communication overhead, such as update time, less representative in reflecting the real-world condition. Therefore, we used communication bits per round as the metric for communication overhead in our study. This metric is solely related to the FL algorithms and remains unchanged when switching between the simulation environment and real-world deployment. Thus, it is a representative and consistent measure of communication complexity of different FL algorithms. In the future, we plan to establish a multi-workstation platform for the comprehensive comparison of GCML with other FL methods in real-world environments.

Additionally, our focus extends to enhancing GCML in several key areas. Firstly, leveraging its inherent mutual learning capability — enabling the simultaneous training of two models with distinct structures [18] — we intend to assess GCML's performance across local sites with heterogeneous model architectures. Moreover, by recognizing the unpredictable real-world scenarios where sites may exit from FL unexpectedly due to factors such as computational facility shutdowns, network connectivity issues [65] or simply because the training takes prolong time due to excessive incoming models, we will rigorously evaluate GCML's robustness to the limited site availability (with site being temporarily unavailable, dropping out or joining during training) in FL processes. Finally, given the paramount importance of data security in FL contexts, we plan to investigate the integration of GCML with established privacy-preserving frameworks like differential privacy [66].

## VII. Conclusion

In this paper, we present a novel FL framework, named GCML, to learn personalized models in a decentralized environment under Gossip Protocol. In this framework, we propose deep contrastive mutual learning to address the challenge of handling heterogeneous incoming models due to non-IID data distributions across different sites. By integrating it with other efforts to encourage each site to optimize its local segmentation model and leverage useful information from the peer, we demonstrated that GCML outperformed the state-of-the-art approaches, in both centralized and decentralized federated learning settings, on three medical image segmentation tasks with significantly reduced communication overhead. With the increase of model size and data scale in modern medical studies, our proposed GCML provides a robust, flexible, and efficient FL solution to simultaneously meet the unique needs of different sites and to potentially facilitate real-world deployment.

## Acknowledgment

This work is supported by a research grant from Varian Medical Systems (Palo Alto, CA, USA), UL1TR001433 from the National Center for Advancing Translational Sciences, and R21EB030209 from the National Institute of Biomedical Imaging and Bioengineering of the National Institutes of Health, National Institutes of Health, USA. The content is solely the responsibility of the authors and does not necessarily represent the official views of the National Institutes of Health. This research has been partially funded through the generous support of Herbert and Florence Irving/the Irving Trust.

TABLE 8
DSCs (%) UNDER DIFFERENT SITE PAIRS

| Site | 1 Pair | 2 Pairs | 4 Pairs |
|---|---|---|---|
| 1 | 85.78 | 79.10 | 88.99 |
| 2 | 79.99 | 83.70 | 84.69 |
| 3 | 95.62 | 95.60 | 94.85 |
| 4 | 94.29 | 94.73 | 94.63 |
| 5 | 94.67 | 94.19 | 93.92 |
| 6 | 86.19 | 90.37 | 91.37 |
| 7 | 88.21 | 88.94 | 89.15 |
| 8 | 94.48 | 94.70 | 94.90 |
| Avg. | **90.69** | **91.18** | **91.87** |

TABLE 9
DSCs (%) UNDER DIFFERENT ROUNDS OF WARMUP

| Site | 0 | 50 | 100 |
|---|---|---|---|
| 1 | 88.99 | 85.74 | 82.34 |
| 2 | 84.69 | 84.15 | 83.93 |
| 3 | 94.85 | 95.26 | 95.18 |
| 4 | 94.63 | 94.18 | 94.20 |
| 5 | 93.92 | 94.57 | 94.19 |
| 6 | 91.37 | 91.97 | 90.94 |
| 7 | 89.15 | 90.19 | 88.67 |
| 8 | 94.90 | 94.55 | 94.12 |
| Avg. | **91.87** | **91.89** | **91.23** |

## REFERENCES

[1] O. Ronneberger, P. Fischer, and T. Brox, "U-Net: Convolutional Networks for Biomedical Image Segmentation," May 2015.

[2] A. Dosovitskiy *et al.*, "An Image is Worth 16x16 Words: Transformers for Image Recognition at Scale," Oct. 2020.

[3] B. McMahan, E. Moore, D. Ramage, S. Hampson, and B. A. y Arcas, "Communication-Efficient Learning of Deep Networks from Decentralized Data," in *Proceedings of the 20th International Conference on Artificial Intelligence and Statistics*, A. Singh and J. Zhu, Eds., in Proceedings of Machine Learning Research, vol. 54. PMLR, Aug. 2017, pp. 1273–1282.

[4] Y. Shen, A. Sowmya, Y. Luo, X. Liang, D. Shen, and J. Ke, "A Federated Learning System for Histopathology Image Analysis With an Orchestral Stain-Normalization GAN," *IEEE Trans Med Imaging*, vol. 42, no. 7, pp. 1969–1981, 2023, doi: 10.1109/TMI.2022.3221724.

[5] S. M. Hosseini, M. Sikaroudi, M. Babaie, and H. R. Tizhoosh, "Proportionally Fair Hospital Collaborations in Federated Learning of Histopathology Images," *IEEE Trans Med Imaging*, vol. 42, no. 7, pp. 1982–1995, 2023, doi: 10.1109/TMI.2023.3234450.

[6] G. Elmas *et al.*, "Federated Learning of Generative Image Priors for MRI Reconstruction," *IEEE Trans Med Imaging*, vol. 42, no. 7, pp. 1996–2009, 2023, doi: 10.1109/TMI.2022.3220757.

[7] C.-M. Feng, Y. Yan, S. Wang, Y. Xu, L. Shao, and H. Fu, "Specificity-Preserving Federated Learning for MR Image Reconstruction," *IEEE Trans Med Imaging*, vol. 42, no. 7, pp. 2010–2021, 2023, doi: 10.1109/TMI.2022.3202106.

[8] X. Gong *et al.*, "Federated Learning With Privacy-Preserving Ensemble Attention Distillation," *IEEE Trans Med Imaging*, vol. 42, no. 7, pp. 2057–2067, 2023, doi: 10.1109/TMI.2022.3213244.

[9] A. Hatamizadeh *et al.*, "Do Gradient Inversion Attacks Make Federated Learning Unsafe?," *IEEE Trans Med Imaging*, vol. 42, no. 7, pp. 2044–2056, 2023, doi: 10.1109/TMI.2023.3239391.

[10] R. Yan *et al.*, "Label-Efficient Self-Supervised Federated Learning for Tackling Data Heterogeneity in Medical Imaging," *IEEE Trans Med Imaging*, vol. 42, no. 7, pp. 1932–1943, 2023, doi: 10.1109/TMI.2022.3233574.

[11] N. Dong, M. Kampffmeyer, I. Voiculescu, and E. Xing, "Federated Partially Supervised Learning With Limited Decentralized Medical Images," *IEEE Trans Med Imaging*, vol. 42, no. 7, pp. 1944–1954, Jul. 2023, doi: 10.1109/TMI.2022.3231017.

[12] S. Pati *et al.*, "Federated learning enables big data for rare cancer boundary detection," *Nat Commun*, vol. 13, no. 1, p. 7346, Dec. 2022, doi: 10.1038/s41467-022-33407-5.

[13] N. Rieke *et al.*, "The future of digital health with federated learning," *NPJ Digit Med*, vol. 3, no. 1, p. 119, Sep. 2020, doi: 10.1038/s41746-020-00323-1.

[14] O. L. Saldanha *et al.*, "Swarm learning for decentralized artificial intelligence in cancer histopathology," *Nat Med*, Jun. 2022, doi: 10.1038/s41591-022-01768-5.

[15] T. Li, S. Hu, A. Beirami, and V. Smith, "Ditto: Fair and Robust Federated Learning Through Personalization," *arXiv: 2012.04221*, Dec. 2020.

[16] J. Wu, W. Bao, E. Ainsworth, and J. He, "Personalized Federated Learning with Parameter Propagation," in *Proceedings of the 29th ACM SIGKDD Conference on Knowledge Discovery and Data Mining*, New York, NY, USA: ACM, Aug. 2023, pp. 2594–2605. doi: 10.1145/3580305.3599464.

[17] A. G. Roy, S. Siddiqui, S. Pölsterl, N. Navab, and C. Wachinger, "BrainTorrent: A Peer-to-Peer Environment for Decentralized Federated Learning," *arXiv:1905.06731*, May 2019.

[18] S. Kalra, J. Wen, J. C. Cresswell, M. Volkovs, and H. R. Tizhoosh, "Decentralized federated learning through proxy model sharing," *Nat Commun*, vol. 14, no. 1, p. 2899, Dec. 2023, doi: 10.1038/s41467-023-38569-4.

[19] C. Li, G. Li, and P. K. Varshney, "Decentralized Federated Learning via Mutual Knowledge Transfer," *IEEE Internet Things J*, vol. 9, no. 2, pp. 1136–1147, Jan. 2022, doi: 10.1109/JIOT.2021.3078543.

[20] Y. Zhang, T. Xiang, T. M. Hospedales, and H. Lu, "Deep Mutual Learning," in *Proceedings of the IEEE Conference on Computer Vision and Pattern Recognition (CVPR)*, 2018, pp. 4320–4328.

[21] S. Pereira, A. Pinto, V. Alves, and C. A. Silva, "Brain Tumor Segmentation Using Convolutional Neural Networks in MRI Images," *IEEE Trans Med Imaging*, vol. 35, no. 5, pp. 1240–1251, May 2016, doi: 10.1109/TMI.2016.2538465.

[22] Y. Yuan, M. Chao, and Y.-C. Lo, "Automatic Skin Lesion Segmentation Using Deep Fully Convolutional Networks With Jaccard Distance," *IEEE Trans Med Imaging*, vol. 36, no. 9, pp. 1876–1886, 2017, doi: 10.1109/TMI.2017.2695227.

[23] N. Shoham *et al.*, "Overcoming Forgetting in Federated Learning on Non-IID Data," *arXiv: 1910.07796*, Oct. 2019.

[24] S. P. Karimireddy, S. Kale, M. Mohri, S. J. Reddi, S. U. Stich, and A. T. Suresh, "SCAFFOLD: Stochastic Controlled Averaging for Federated Learning," Oct. 2019.

[25] T. Li, A. K. Sahu, M. Zaheer, M. Sanjabi, A. Talwalkar, and V. Smith, "Federated Optimization in Heterogeneous Networks," Dec. 2018.

[26] L. Mächler, I. Ezhov, S. Shit, and J. C. Paetzold, "FedPIDAvg: A PID Controller Inspired Aggregation Method for Federated Learning," in *Brainlesion: Glioma, Multiple Sclerosis, Stroke and Traumatic Brain Injuries*, 2023, pp. 209–217. doi: 10.1007/978-3-031-44153-0_20.

[27] E. Isik-Polat, G. Polat, A. Kocyigit, and A. Temizel, "Evaluation and Analysis of Different Aggregation and Hyperparameter Selection Methods for Federated Brain Tumor Segmentation," in *Brainlesion: Glioma, Multiple Sclerosis, Stroke and Traumatic Brain Injuries*, 2022, pp. 405–419. doi: 10.1007/978-3-031-09002-8_36.

[28] A. Hashemi, A. Acharya, R. Das, H. Vikalo, S. Sanghavi, and I. S. Dhillon, "On the Benefits of Multiple Gossip Steps in Communication-Constrained Decentralized Federated Learning," *IEEE Transactions on Parallel and Distributed Systems*, pp. 1–1, 2022, doi: 10.1109/TPDS.2021.3138977.

[29] L. He, S. P. Karimireddy, and M. Jaggi, "Byzantine-Robust Decentralized Learning via ClippedGossip," Feb. 2022.

[30] C. Che, X. Li, C. Chen, X. He, and Z. Zheng, "A Decentralized Federated Learning Framework via Committee Mechanism With Convergence Guarantee," *IEEE Transactions on Parallel and Distributed Systems*, vol. 33, no. 12, pp. 4783–4800, Dec. 2022, doi: 10.1109/TPDS.2022.3202887.

[31] Z. Gao, F. Wu, W. Gao, and X. Zhuang, "A New Framework of Swarm Learning Consolidating Knowledge From Multi-Center Non-IID Data for Medical Image Segmentation," *IEEE Trans Med Imaging*, vol. 42, no. 7, pp. 2118–2129, Jul. 2023, doi: 10.1109/TMI.2022.3220750.

[32] L. Collins, H. Hassani, A. Mokhtari, and S. Shakkottai, "Exploiting Shared Representations for Personalized Federated Learning," in *Proceedings of the 38th International Conference on Machine Learning*, M. Meila and T. Zhang, Eds., in Proceedings of Machine Learning Research, vol. 139. PMLR, Mar. 2021, pp. 2089–2099. [Online]. Available: https://proceedings.mlr.press/v139/collins21a.html

[33] Z. Chen, M. Zhu, C. Yang, and Y. Yuan, "Personalized Retrogress-Resilient Framework for Real-World Medical Federated Learning," 2021, pp. 347–356. doi: 10.1007/978-3-030-87199-4_33.

[34] M. Jiang, H. Yang, C. Cheng, and Q. Dou, "IOP-FL: Inside-Outside Personalization for Federated Medical Image Segmentation," *IEEE Trans Med Imaging*, vol. 42, no. 7, pp. 2106–2117, Jul. 2023, doi: 10.1109/TMI.2023.3263072.

[35] R. Baraglia, P. Dazzi, M. Mordacchini, and L. Ricci, "A peer-to-peer recommender system for self-emerging user communities based on gossip overlays," *J Comput Syst Sci*, vol. 79, no. 2, pp. 291–308, Mar. 2013, doi: 10.1016/j.jcss.2012.05.011.

[36] I. Hegedűs, G. Danner, and M. Jelasity, "Gossip Learning as a Decentralized Alternative to Federated Learning," 2019, pp. 74–90. doi: 10.1007/978-3-030-22496-7_5.

[37] L. Yuan, Y. Ma, L. Su, and Z. Wang, "Peer-to-Peer Federated Continual Learning for Naturalistic Driving Action Recognition," in *2023 IEEE/CVF Conference on Computer Vision and Pattern Recognition Workshops (CVPRW)*, IEEE, Jun. 2023, pp. 5250–5259. doi: 10.1109/CVPRW59228.2023.00553.

[38] T. Shen *et al.*, "Federated Mutual Learning," *arXiv: 2006.16765*, Jun. 2020.

[39] R. Yang, J. Tian, and Y. Zhang, "Regularized Mutual Learning for Personalized Federated Learning," in *Proceedings of The 13th Asian Conference on Machine Learning*, V. N. Balasubramanian and I. Tsang, Eds., in Proceedings of Machine Learning Research, vol. 157. PMLR, Sep. 2021, pp. 1521–1536. [Online]. Available: https://proceedings.mlr.press/v157/yang21c.html

[40] W. Huang, M. Ye, and B. Du, "Learn From Others and Be Yourself in Heterogeneous Federated Learning," in *Proceedings of the IEEE/CVF Conference on Computer Vision and Pattern Recognition (CVPR)*, Jun. 2022, pp. 10143–10153.

[41] C. Wu, F. Wu, L. Lyu, Y. Huang, and X. Xie, "Communication-efficient federated learning via knowledge distillation," *Nat Commun*, vol. 13, no. 1, p. 2032, Apr. 2022, doi: 10.1038/s41467-022-29763-x.

[42] C. Wu, F. Wu, L. Lyu, Y. Huang, and X. Xie, "FedKD: Communication Efficient Federated Learning via

Knowledge Distillation," *arXiv: 2108.13323*, Aug. 2021, doi: 10.1038/s41467-022-29763-x.

[43] Y. Yuan, "Automatic Brain Tumor Segmentation with Scale Attention Network," in *Brainlesion: Glioma, Multiple Sclerosis, Stroke and Traumatic Brain Injuries*, A. Crimi and S. Bakas, Eds., Cham: Springer International Publishing, 2021, pp. 285–294.

[44] Y. Yuan, "Automatic Head and Neck Tumor Segmentation in PET/CT with Scale Attention Network," in *Head and Neck Tumor Segmentation*, V. Andrearczyk, V. Oreiller, and A. Depeursinge, Eds., Cham: Springer International Publishing, 2021, pp. 44–52.

[45] V. Oreiller *et al.*, "Head and neck tumor segmentation in PET/CT: The HECKTOR challenge," *Med Image Anal*, vol. 77, p. 102336, Apr. 2022, doi: 10.1016/j.media.2021.102336.

[46] Y. Yuan, "Evaluating Scale Attention Network for Automatic Brain Tumor Segmentation with Large Multi-parametric MRI Database," in *International MICCAI Brainlesion workshop*, 2022, pp. 42–53. doi: 10.1007/978-3-031-09002-8_4.

[47] K. He, X. Zhang, S. Ren, and J. Sun, "Deep Residual Learning for Image Recognition," in *2016 IEEE Conference on Computer Vision and Pattern Recognition (CVPR)*, IEEE, Jun. 2016, pp. 770–778. doi: 10.1109/CVPR.2016.90.

[48] U. Baid *et al.*, "The RSNA-ASNR-MICCAI BraTS 2021 Benchmark on Brain Tumor Segmentation and Radiogenomic Classification," Jul. 2021.

[49] V. Andrearczyk *et al.*, "Overview of the HECKTOR Challenge at MICCAI 2021: Automatic Head and Neck Tumor Segmentation and Outcome Prediction in PET/CT Images," in *Head and Neck Tumor Segmentation and Outcome Prediction*, V. Andrearczyk, V. Oreiller, M. Hatt, and A. Depeursinge, Eds., Cham: Springer International Publishing, 2022, pp. 1–37.

[50] Z. Zhang *et al.*, "Large-Scale Multi-Center CT and MRI Segmentation of Pancreas with Deep Learning," *arXiv: 2405.12367*, May 2024.

[51] M. Manthe, S. Duffner, and C. Lartizien, "Federated brain tumor segmentation: An extensive benchmark," *Med Image Anal*, vol. 97, p. 103270, Oct. 2024, doi: 10.1016/j.media.2024.103270.

[52] S. S. Shapiro and M. B. Wilk, "An Analysis of Variance Test for Normality (Complete Samples)," *Biometrika*, vol. 52, no. 3/4, p. 591, Dec. 1965, doi: 10.2307/2333709.

[53] H. B. Mann and D. R. Whitney, "On a Test of Whether one of Two Random Variables is Stochastically Larger than the Other," *The Annals of Mathematical Statistics*, vol. 18, no. 1, pp. 50–60, Mar. 1947, doi: 10.1214/aoms/1177730491.

[54] F. Isensee, P. F. Jaeger, S. A. A. Kohl, J. Petersen, and K. H. Maier-Hein, "nnU-Net: a self-configuring method for deep learning-based biomedical image segmentation," *Nat Methods*, vol. 18, no. 2, pp. 203–211, Feb. 2021, doi: 10.1038/s41592-020-01008-z.

[55] L. van der Maaten and G. E. Hinton, "Visualizing Data using t-SNE," *Journal of Machine Learning Research*, vol. 9, pp. 2579–2605, 2008, [Online]. Available: https://api.semanticscholar.org/CorpusID:5855042

[56] A. R. Jamieson, M. L. Giger, K. Drukker, H. Li, Y. Yuan, and N. Bhooshan, "Exploring nonlinear feature space dimension reduction and data representation in breast CADx with Laplacian eigenmaps and -SNE," *Med Phys*, vol. 37, no. 1, pp. 339–351, Jan. 2010, doi: 10.1118/1.3267037.

[57] T. Chen, S. Kornblith, M. Norouzi, and G. Hinton, "A simple framework for contrastive learning of visual representations," in *Proceedings of the 37th International Conference on Machine Learning*, in ICML'20. JMLR.org, 2020.

[58] K. He, H. Fan, Y. Wu, S. Xie, and R. Girshick, "Momentum contrast for unsupervised visual representation learning," in *Proceedings of the IEEE/CVF conference on computer vision and pattern recognition*, 2020, pp. 9729–9738.

[59] P. Khosla *et al.*, "Supervised contrastive learning," *Adv Neural Inf Process Syst*, vol. 33, pp. 18661–18673, 2020.

[60] Q. Li, B. He, and D. Song, "Model-Contrastive Federated Learning," in *2021 IEEE/CVF Conference on Computer Vision and Pattern Recognition (CVPR)*, IEEE, Jun. 2021, pp. 10708–10717. doi: 10.1109/CVPR46437.2021.01057.

[61] X. Xu and S.-L. Huang, "On Distributed Learning With Constant Communication Bits," *IEEE Journal on Selected Areas in Information Theory*, vol. 3, no. 1, pp. 125–134, Mar. 2022, doi: 10.1109/JSAIT.2022.3157797.

[62] P. Kairouz *et al.*, "Advances and Open Problems in Federated Learning," *Foundations and Trends® in Machine Learning*, vol. 14, no. 1–2, pp. 1–210, 2021, doi: 10.1561/2200000083.

[63] J. Zhao, R. Li, H. Wang, and Z. Xu, "HotFed: Hot Start through Self-Supervised Learning in Federated Learning," in *2021 IEEE 23rd Int Conf on High*

*Performance Computing & Communications; 7th Int Conf on Data Science & Systems; 19th Int Conf on Smart City; 7th Int Conf on Dependability in Sensor, Cloud & Big Data Systems & Application (HPCC/DSS/SmartCity/DependSys)*, IEEE, Dec. 2021, pp. 149–156. doi: 10.1109/HPCC-DSS-SmartCity-DependSys53884.2021.00046.

[64] Y. Zhao, M. Li, L. Lai, N. Suda, D. Civin, and V. Chandra, "Federated Learning with Non-IID Data," Jun. 2018, doi: 10.48550/arXiv.1806.00582.

[65] H. Wang and J. Xu, "Combating Client Dropout in Federated Learning via Friend Model Substitution," May 2022.

[66] C. Dwork, F. McSherry, K. Nissim, and A. Smith, "Calibrating Noise to Sensitivity in Private Data Analysis," 2006, pp. 265–284. doi: 10.1007/11681878_14.